\documentclass[%
 reprint,
 amsmath,amssymb,
 aps,
pra,
]{revtex4-1}

\usepackage{graphicx}
\usepackage{dcolumn}
\usepackage{bm}

\usepackage{braket} 

\newcommand{\Ortho}{\mathrm{O}}
\newcommand{\SO}{\mathrm{SO}}
\newcommand{\U}{\mathrm{U}}
\newcommand{\SU}{\mathrm{SU}}

\begin{document}

\title{Asymptotic Limits of the Wigner $12J$-Symbol \\ In Terms of the Ponzano-Regge Phases}

\author{Liang Yu}%
\email{liangyu@wigner.berkeley.edu}
\affiliation{%
 Department of Physics, University of California, Berkeley, California 94720 USA \\
}%

\date{\today}

\begin{abstract}

There are two types of asymptotic formulas for the $12j$ symbol with one small and 11 large angular momenta. We have derived the first type of formula previously in [L.\ Yu, Phys.\ Rev.\ A{\bf 84} 022101 (2011)]. We will derive the second type in this paper. We find that this second asymptotic formula for the $12j$ symbol is expressed in terms of the vector diagram associated with two $6j$ symbols, namely, the vector diagram of two adjacent tetrahedra sharing a common face. As a result, two sets of Ponzano-Regge phases appear in the asymptotic formula. This work contributes another asymptotic formula of the Wigner $12j$ symbol to the re-coupling theory of angular momenta. 

\begin{description}
\item[PACS numbers] 03.65.Sq, 02.30.Ik, 03.65.Vf
\end{description}
\end{abstract}

\pacs{03.65.Sq, 02.30.Ik, 03.65.Vf}
                             
\maketitle

\section{Introduction}

The Wigner $12j$ symbol is described in various textbooks on angular momentum theory \cite{edmonds1960, brink1968, biedenharn1981, yutsis1962}. It has applications in atomic physics \cite{thole1988, ceulemans1993} and loop quantum gravity \cite{carfora1993}. In this paper, we derive an asymptotic formula for the $12j$ symbol in the limit of one small and 11 large angular momenta. 

There are two special formulas for the exact $12j$ symbol when one of its 12 arguments is zero \cite{jahn1954}. These special formulas are displayed in Eq.\ (\ref{eq_jahn_A9}) and Eq.\ (\ref{eq_jahn_A8}). In an earlier paper \cite{yu2011b}, we derived an asymptotic formula of the $12j$ symbol where the zero parameter in Eq.\ (\ref{eq_jahn_A9}) is replaced by a small parameter, and the other 11 parameters are taken to be large. In this paper, we will derive an asymptotic formula for the $12j$ symbol where the zero parameter in Eq.\ (\ref{eq_jahn_A8}) is replaced by a small parameter, and the other 11 parameters are taken to be large. 

The main theoretical tool we use is a generalization of the Born-Oppenheimer approximation, called multicomponent WKB theory \cite{littlejohn1991, littlejohn1992, weinstein1996, yu2011a, yu2011b}, in which the small angular momenta are modeled by exact linear algebra, and the large angular momenta are modeled by a WKB wave function. Each wave function in this model consists of a spinor factor and a factor in the form of a scalar WKB solution. A gauge-invariant expression for the resulting multicomponent WKB wave function is developed in the semiclassical analysis of the $9j$ symbol with small and large angular momenta \cite{yu2011a}. This gauge-invariant expression plays a crucial role in deriving the results in \cite{yu2011a, yu2011b}, as well as the result in this paper. Thus, this paper assumes familiarity with it. 

In our earlier paper \cite{yu2011b}, we find that the first type of asymptotic formula for the $12j$ symbol  is based on the geometry associated with the $9j$ symbol on the right hand side of Eq.\ (\ref{eq_jahn_A9}). Thus, we expect the second type of asymptotic formulas for the $12j$ symbol with a small angular momentum to be expressed in terms of geometries associated with the two $6j$ symbols on the right hand side of Eq.\ (\ref{eq_jahn_A8}). This is in fact the case. The asymptotic formula of the $6j$ symbol in terms of the Ponzano-Regge phase \cite{ponzano-regge-1968} is well known from the role it plays in Regge gravity \cite{regge1961} and topological quantum field theory \cite{ooguri1991, ooguri1992}. Using this asymptotic formula for the $6j$ symbol, we find that the second type of asymptotic formula for the $12j$ symbol contains two Ponzano-Regge phases. 

We will now give an outline of this paper. In Sec.\ \ref{sec_12j_definition}, we display two special formulas for the exact $12j$ symbol, and then express the $12j$ symbol as an inner product between two multicomponent wave functions. In Sec.\ \ref{sec_wave functions}, we use the procedure outlined in \cite{yu2011a} to find the gauge-invariant multicomponent WKB form of these wave functions. In Sec.\ \ref{sec_lagrangian_manifolds}, based on the methods developed in \cite{littlejohn2010b}, we sketch the semiclassical analysis of these Lagrangian manifolds and use the Ponzano Regge formula to confirm the result of the analysis. In Sec.\ \ref{sec_spinor_product}, we find the spinor inner products at the intersections of the Lagrangian manifolds. Putting it all together, we derive an asymptotic formula for the $12j$ symbol in Sec.\ \ref{sec_final_formula}. The last section contains comments and discussions.

\section{\label{sec_12j_definition}The $12j$-Symbol}

The $12j$ symbol was first defined by Jahn and Hope \cite{jahn1954} in 1954. The appendix of their paper gives two special formulas for the $12j$ symbol when one of its 12 parameters is zero. In the following, we rewrite Eq.\ (A9) and Eq.\ (A8) in that appendix using a more convenient labeling. 

\begin{widetext}
\begin{eqnarray}
\label{eq_jahn_A9}
\left\{
   \begin{array}{cccc}
    j_1 & j_{2} & j_{12} & j_{12} \\ 
    j_{3} & j_4 & j_{34} &  j_{13}  \\ 
    j_{13} & j_{24} & 0 & j_6  \\
  \end{array} 
  \right \}  
=
\left\{
   \begin{array}{cccc}
    0 & j_{13} & j_{13} & j_{3} \\ 
    j_{12} & j_6 & j_{34} &  j_{2}  \\ 
    j_{12} & j_{24} & j_1 & j_4  \\
  \end{array} 
  \right \}    
&=&
\left\{
   \begin{array}{cccc}
    j_6 & j_{12} & j_{34} & j_{3} \\ 
    j_{13} & 0 & j_{13} &  j_{2}  \\ 
    j_{24} & j_{12} & j_4 & j_1  \\
  \end{array} 
  \right \}  
=
\left\{
   \begin{array}{cccc}
    j_4 & j_{2} & j_{24} & j_{13} \\ 
    j_{3} & j_1 & j_{13} &  j_{12}  \\ 
    j_{34} & j_{12} & j_6 & 0  \\
  \end{array} 
  \right \}   \nonumber  \\
&=&
\frac{1}{\sqrt{[j_{12}] \, [j_{13}]}} \, 
\left\{
   \begin{array}{ccc}
    j_1 & j_{2} & j_{12}  \\ 
    j_{3} & j_4 & j_{34}   \\ 
    j_{13} & j_{24} & j_6   \\
  \end{array} 
  \right \}  
\end{eqnarray}

\begin{eqnarray}
\label{eq_jahn_A8}
\left\{
   \begin{array}{cccc}
    j_1 & 0 & j_{1} & j_{346} \\ 
    j_3 & j_4 & j_{34} &  j_{135}  \\ 
    j_{13} & j_{4} & j_5 & j_6  \\
  \end{array} 
  \right \}    
&=&
\left\{
   \begin{array}{cccc}
    j_4 & j_4 & 0 & j_{1} \\ 
    j_{34} & j_6 & j_{346} &  j_{13}  \\ 
    j_{3} & j_{135} & j_1 & j_5  \\
  \end{array} 
  \right \}  
=
\left\{
   \begin{array}{cccc}
    j_5 & j_{346} & j_{1} & 0 \\ 
    j_{135} & j_6 & j_{4} &  j_{3}  \\ 
    j_{13} & j_{34} & j_1 & j_4  \\
  \end{array} 
  \right \}  
=
\left\{
   \begin{array}{cccc}
    j_1 & j_{3} & j_{13} & j_{135} \\ 
    0 & j_4 & j_{4} &  j_{346}  \\ 
    j_{1} & j_{34} & j_5 & j_6  \\
  \end{array} 
  \right \}   \nonumber  \\
  \left\{
   \begin{array}{cccc}
    j_5 & j_{13} & j_{135} & j_{4} \\ 
    j_1 & j_1 & 0 &  j_{34}  \\ 
    j_{346} & j_{3} & j_6 & j_4  \\
  \end{array} 
  \right \}    
&=&
\left\{
   \begin{array}{cccc}
    j_5 & j_{135} & j_{13} & j_{3} \\ 
    j_{346} & j_6 & j_{34} &  0  \\ 
    j_{1} & j_{4} & j_1 & j_4  \\
  \end{array} 
  \right \}  
=
\left\{
   \begin{array}{cccc}
    j_4 & j_{34} & j_{3} & j_{13} \\ 
    j_{4} & j_6 & j_{135} &  j_{1}  \\ 
    0 & j_{346} & j_1 & j_5  \\
  \end{array} 
  \right \}  
=
\left\{
   \begin{array}{cccc}
    j_5 & j_{1} & j_{346} & j_{34} \\ 
    j_{13} & j_1 & j_{3} &  j_{4}  \\ 
    j_{135} & 0 & j_6 & j_4  \\
  \end{array} 
  \right \}   \nonumber  \\
&=&
\frac{(-1)^{j_1+2j_3 + j_4 + j_{346} + j_{135} + j_5 + j_6}}{\sqrt{[j_1]\, [j_4]}} \, 
\left\{
   \begin{array}{ccc}
    j_{346} & j_3 & j_{135}  \\ 
    j_4 & j_6 & j_{34}   \\ 
  \end{array} 
  \right \}  \, 
\left\{
   \begin{array}{ccc}
    j_{346} & j_3 & j_{135}  \\ 
    j_{13}& j_5 & j_{1}   \\ 
  \end{array} 
  \right \}  
\end{eqnarray}
\end{widetext}
where the square bracket notation $[c]$ stands for $[c] = 2c+1$. In this paper, we will replace the zero parameter in Eq.\ (\ref{eq_jahn_A8}) by a nonzero small parameter $j_2 = s_2$, and take the other 11 parameters to be large compared to $s_2$.

From Eq.\ (A4) in \cite{jahn1954}, the $12j$ symbol is defined as a scalar product

\begin{equation}
\label{eq_12j_definition}
\left\{
   \begin{array}{cccc}
    j_1 & s_2 & j_{12} & j_{346} \\ 
    j_3 & j_4 & j_{34} &  j_{135}  \\ 
    j_{13} & j_{24} & j_5 & j_6  \\
  \end{array} 
  \right \}    
= \frac{\braket{ b  |  a } }{ \{ [j_{12}][j_{34}][j_{13}][j_{24}] [j_{346}] [j_{135}] \}^{\frac{1}{2}}} 
\end{equation}
where the square bracket notation $[\cdot]$ again denotes $[c] = 2c + 1$, and $\ket{a}$ and $\ket{b}$ are normalized simultaneous eigenstates of lists of operators with certain eigenvalues. We will ignore the phase conventions of $\ket{a}$ and $\ket{b}$ for now, since we did not use them to derive our formula. In our notation, the two states are

\begin{equation}
\label{eq_a_state}
\ket{a} =  \left| 
\begin{array} { @{\,}c@{\,}c@{\,}c@{\,}c@{\;}c@{\;}c@{\;}c@{\;}c@{\;}c@{\;}c@{}}
	\hat{I}_1 &  {\bf S}_2^2  & \hat{I}_3  & \hat{I}_4 & \hat{I}_5 & \hat{I}_6 & \hat{\bf J}_{12}^2 & \hat{\bf J}_{34}^2 & \hat{\bf J}_{346}^2 &  \hat{\bf J}_{\text{tot}}  \\
	j_1 & s_2 & j_3 & j_4 & j_5 & j_6 & j_{12} & j_{34} & j_{346} & {\bf 0} 
\end{array}  \right>
\end{equation}

\begin{equation}
\label{eq_b_state}
\ket{b} =  \left| 
\begin{array} { @{\,}c@{\,}c@{\,}c@{\,}c@{\;}c@{\;}c@{\;}c@{\;}c@{\;}c@{\;}c@{}}
	\hat{I}_1 &  {\bf S}_2^2  & \hat{I}_3  & \hat{I}_4 & \hat{I}_5 & \hat{I}_6 & \hat{\bf J}_{13}^2 & \hat{\bf J}_{24}^2 & \hat{\bf J}_{135}^2 &  \hat{\bf J}_{\text{tot}}  \\
	j_1 & s_2 & j_3 & j_4 & j_5 & j_6 & j_{13} & j_{24} & j_{135} & {\bf 0} 
\end{array}  \right>
\end{equation}
In the above notation, the large ket lists the operators on the top row, and the corresponding quantum numbers are listed on the bottom row. The hat is used to distinguish differential operators from their symbols, that is, the associated classical functions. 

The states $\ket{a}$ and $\ket{b}$ live in a total Hilbert space of six angular momenta ${\mathcal H}_1 \otimes {\mathcal H}_3 \otimes {\mathcal H}_4 \otimes {\mathcal H}_5 \otimes {\mathcal H}_6 \otimes {\mathcal H}_{s}$, where $s= s_2$. Each large angular momentum ${\bf J}_r$, $r = 1,3,4,5,6$, is represented by a Schwinger Hilbert space of two harmonic oscillators, namely, ${\bf H}_r = L^2 ({\mathbb R}^2)$ \cite{littlejohn2007}. The small angular momentum ${\bf S}$ is represented by the usual $2s+1$ dimensional representation of $\SU(2)$, that is, ${\mathcal H}_s = {\mathbb C}^{2s+1}$.

Let us now define the lists of operators in (\ref{eq_a_state}) and (\ref{eq_b_state}). First we look at the operators $\hat{I}_r$, $r = 1,3,4,5,6$, and ${\bf J}_{34}^2$, ${\bf J}_{346}^2$, ${\bf J}_{13}^2$, ${\bf J}_{135}^2$, which act only on the large angular momentum spaces ${\mathcal H}_r$, each of which can be viewed as a space of wave functions $\psi(x_{r1}, x_{r2})$ for two harmonic oscillators of unit frequency and mass. Let $\hat{a}_{r\mu} = (\hat{x}_{r\mu} + i \hat{p}_{r\mu})/\sqrt{2}$ and $\hat{a}_{r\mu}^\dagger = (\hat{x}_{r\mu} - i \hat{p}_{r\mu})/\sqrt{2}$, $\mu = 1,2$, be the usual annihilation and creation operators. The operators $\hat{I}_r$ and $\hat{J}_{ri}$ are constructed from these differential operators $\hat{a}$ and $\hat{a}^\dagger$ as follows, 

\begin{equation}
\hat{I}_r = \frac{1}{2} \, \hat{a}_r^\dagger \hat{a}_r \, ,  \quad \quad \hat{J}_{ri} = \frac{1}{2} \, \hat{a}^\dagger_r \sigma_i \hat{a}_r  \, ,
\end{equation}
where $i = 1,2,3$, and $\sigma_i$ are the Pauli matrices. The quantum numbers $j_r$, $r = 1,3,4,5,6$ specify the eigenvalues of both $\hat{I}_r$ and $\hat{\bf J}_r^2$, to be $j_r$ and $j_r(j_r + 1)$, respectively.

The operators $\hat{\bf J}_{34}^2$, $\hat{\bf J}_{346}^2$, $\hat{\bf J}_{13}^2$, $\hat{\bf J}_{135}^2$ that define intermediate coupling of the large angular momenta are defined by partial sums of $\hat{\bf J}_r$,

\begin{equation}
\label{eq_J34_J346_vector}
	\hat{\bf J}_{34} = \hat{\bf J}_3 + \hat{\bf J}_4 \, ,   \quad \quad  \hat{\bf J}_{346} = \hat{\bf J}_3 + \hat{\bf J}_4 + \hat{\bf J}_6 \, .
\end{equation}

\begin{equation}
\label{eq_J13_J135_vector}
	\hat{\bf J}_{13} = \hat{\bf J}_1 + \hat{\bf J}_3 \, ,   \quad \quad  \hat{\bf J}_{135} = \hat{\bf J}_1 + \hat{\bf J}_3 + \hat{\bf J}_5  \, .
\end{equation}

The quantum numbers $j_{i}$ , $i = 34, 346, 13, 135$ specify the eigenvalues of the operators $\hat{\bf J}_{i}^2$  to be $j_{i} (j_{i}+1)$, for $i = 34, 346, 13, 135$. See  \cite{littlejohn2007} for more detail on the Schwinger model.

Now we turn our attention to the operator $S^2$ that act only on the small angular momentum space ${\mathbb C}^{2s+1}$. Let ${\bf S}$ be the vector of dimensionless spin operators represented by $2s+1$ dimensional matrices that satisfy the $\SU(2)$ commutation relations

\begin{equation}
[S_i, S_j] = i \, \epsilon_{ijk} \, S_k \, .
\end{equation}
The Casimir operator, ${\bf S}^2 = s (s+1)$, is proportional to the identity operator, so its eigenvalue equation is trivially satisfied.

The remaining operators ${\bf \hat{J}}_{12}^2$, ${\bf \hat{J}}_{24}^2$, and ${\bf \hat{J}}_{\text{tot}}$ are non-diagonal matrices of differential operators. They are defined in terms of the operators $\hat{I}_r$, $\hat{\bf J}_{ri}$, and ${\bf S}_i$ as follows,

\begin{eqnarray}
\label{eq_J12_square}
({\hat{J}}_{12}^2)_{\alpha \beta} &=& [  J_1^2+ \hbar^2 s(s+1) ] \delta_{\alpha \beta} + 2 {\bf \hat{J}}_{1}  \cdot {\bf S}_{\alpha \beta} , \quad \quad  \\
\label{eq_J24_square}
({\hat{J}}_{24}^2)_{\alpha \beta}  &=& [  J_{4}^2 + \hbar^2 s(s+1) ] \delta_{\alpha \beta} + 2 {\bf \hat{J}}_{4}  \cdot {\bf S}_{\alpha \beta} ,   \\
\label{eq_Jtot_vector}
({\bf \hat{J}}_{\text{tot}})_{\alpha \beta}  &=& ( {\bf \hat{J}}_1  + {\bf \hat{J}}_3 + {\bf \hat{J}}_4 + {\bf \hat{J}}_5 + {\bf \hat{J}}_6 ) \delta_{\alpha \beta} + \hbar \, {\bf S}_{\alpha \beta} .
\end{eqnarray}
These three operators act nontrivially on both the large and  small angular momentum Hilbert spaces.

\section{\label{sec_wave functions}Multicomponent WKB Wavefunctions}

We follow the approach used in \cite{yu2011a} to find a gauge-invariant form of the multicomponent wave functions $\psi_{\alpha}^a (x) = \braket{x, \alpha | a}$ and $\psi_\alpha^b(x) = \braket{x, \alpha | b}$. Let us focus on $\psi_\alpha^a(x)$, since the treatment for $\psi^b$ is analogous. We will  drop the index $a$ for now.

Let $\hat{D}_i$, $i = 1, \dots, 12$ denote the the operators listed in the definition of $\ket{a}$ in Eq.\  (\ref{eq_a_state}). We seek a unitary operator $\hat{U}$, such that $\hat{D}_i$ for all $i=1, \dots, 12$ are diagonalized when conjugated by $\hat{U}$. In other words,   

\begin{equation}
\hat{U}^\dagger_{\alpha \, \mu} (\hat{D}_i)_{\alpha \, \beta} \, \hat{U}_{\beta \, \nu} = (\hat{\Lambda}_i)_{\mu \, \nu} \, , 
\end{equation}
where $\hat{\Lambda}_i$, $i =1, \dots, 12$ is a list of diagonal matrix operators. Let $\phi^{(\mu)}$ be the simultaneous eigenfunction for the $\mu^{\text{th}}$ diagonal entries $\hat{\lambda}_i$ of the operators $\hat{\Lambda}_i$, $i = 1, \dots, 12$. Then we obtain a simultaneous eigenfunction $\psi_\alpha^{(\mu)}$ of the original list of operators $\hat{D}_i$ from

\begin{equation}
\psi_\alpha^{(\mu)} = \hat{U}_{\alpha \, \mu} \, \phi^{(\mu)} \, . 
\end{equation}
Since we are interested in $\psi_\alpha$ only to first order in $\hbar$, all we need are the zeroth order Weyl symbol matrix $U$ of $\hat{U}$, and the first order symbol matrix $\Lambda_i$ of $\hat{\Lambda}_i$. The resulting asymptotic form of the wave function $\psi(x)$ is a product of a scalar WKB part $B e^{iS}$ and a spinor part $\tau$, that is,

\begin{equation}
\label{eq_general_wave function}
\psi_\alpha^{(\mu)} (x) =  B(x) \, e^{i \, S(x) / \hbar}  \, \tau_{\alpha}^{(\mu)} (x, p) \, . 
\end{equation}
Here the action $S(x)$ and the amplitude $B(x)$ are simultaneous solutions to the Hamilton-Jacobi and the transport equations, respectively, that are associated with the Hamiltonians $\lambda^{(\mu)}_i$. The spinor $\tau^{\mu}$ is the $\mu^{\text{th}}$ column of the matrix $U$, 

\begin{equation}
\label{eq_U_and_tau}
\tau_{\alpha}^{(\mu)} (x, p) = U_{\alpha \mu} (x, p) \, , 
\end{equation}
where $p = \partial S(x) / \partial x$. 

The Weyl symbols of the operators  $\hat{I}_r$ and $\hat{J}_{ri}$, $r = 1,3,4,5,6$, are $I_r - 1/2$ and $J_{ri}$, respectively, where 

\begin{equation}
\label{symbol_I_J}
I_r = \frac{1}{2} \,  \sum_\mu \overline{z}_{r\mu} z_{r\mu} \, ,   \quad \quad J_{ri} = \frac{1}{2} \, \sum_{\mu\nu} \overline{z}_{r\mu} (\sigma^i)_{\mu\nu} z_{r\nu} \, ,
\end{equation}
and where  $z_{r\mu} = x_{r\mu} + ip_{r\mu}$ and $\overline{z}_{r\mu} = x_{r\mu} - ip_{r\mu}$ are the symbols of $\hat{a}$ and $\hat{a}^\dagger$, respectively. The symbols of the remaining operators have the same expressions as Eqs.\  (\ref{eq_J34_J346_vector}), (\ref{eq_J13_J135_vector}), (\ref{eq_J12_square})-(\ref{eq_Jtot_vector}), but without the hats.

Among the operators $\hat{D}_i$, $\hat{J}_{12}^2$ and the vector of the three operators $\hat{\bf J}_{\rm tot}$ are non-diagonal. By looking at (\ref{eq_J12_square}), the expression for $\hat{J}_{12}^2$,  we see that the zeroth order term of the symbol matrix $J_{12}^2$ is already proportional to the identity matrix, so the spinor  $\tau$ must be an eigenvector for the first order term ${\bf J}_{1} \cdot {\bf S}$. Let $\tau^{(\mu)}({\bf J}_{1})$ be the eigenvector of the matrix ${\bf J}_{1} \cdot {\bf S}$ with eigenvalue $\mu J_{1}$, that is, it satisfies

\begin{equation}
\label{eq_JS_eigenvector}
({\bf J}_{1} \cdot {\bf S})_{\alpha \beta} \, \tau^{(\mu)}_\beta = \mu J_{1} \, \tau^{(\mu)}_\beta \, ,
\end{equation}
where $\mu = -s,\, \dots\, , \,+s$. In order to preserve the diagonal symbol matrices $J_{1}$ through the unitary transformation, we must choose the spinor $\tau^{(\mu)}$ to depend only on the direction of ${\bf J}_{1}$. One possible choice of $\tau^{(\mu)}$ is the north standard gauge, (see Appendix A of \cite{littlejohn1992}), in which the spinor $\delta_{\alpha\, \mu}$ is rotated along a great circle from the $z$-axis to the direction of ${\bf J}_{1}$. Explicitly, 

\begin{equation}
\label{tau_north_standard_gauge}
\tau^{(\mu)}_\alpha ({\bf J}_{1}) = e^{i (\mu - \alpha) \phi_{1}} \, d^{(s)}_{\alpha \, \mu} (\theta_{1}) \, , 
\end{equation}
where $(\theta_{1}, \phi_{1})$ are the spherical coordinates that specify the direction of ${\bf J}_{1}$.  Note that this is not the only choice, since Eq.\  (\ref{eq_JS_eigenvector}) is invariant under a local $\U(1)$ gauge transformations.  In other words, any other spinor $\tau' = e^{i g({\bf J}_{1})} \, \tau$ that is related to $\tau$ by a $\U(1)$ gauge transformation satisfies Eq.\  (\ref{eq_JS_eigenvector}).  This local gauge freedom is parametrized by the vector potential,

\begin{equation}
\label{def_A_vec}
{\bf A}^{(\mu)}_{1} = i (\tau^{(\mu)} )^\dagger \, \frac{\partial \tau^{(\mu)}}{\partial {\bf J}_{1}} \, , 
\end{equation}
which transforms as ${\bf A}^{(\mu)'} = {\bf A}^{(\mu)} - \nabla_{{\bf J}_{1}} (g)$ under a local gauge transformation. Moreover, the gradient of the spinor can be expressed in terms of the vector potential, (Eq.\ (A.22) in \cite{littlejohn1992}), as follows, 

\begin{equation}
\label{eq_tau_derivative}
\frac{\partial \tau^{(\mu)}}{\partial {\bf J}_{1}} = i \left( - {\bf A}_{1}^{(\mu)} + \frac{{\bf J}_{1} \times {\bf S}}{J_{1}^2} \right) \, \tau^{(\mu)} \, .
\end{equation}
Once we obtain the complete set of spinors $\tau^{(\mu)}$, $\mu = -s, \dots, s$, we can construct the zeroth order symbol matrix $U$ of the unitary transformation $\hat{U}$ from Eq.\  (\ref{eq_U_and_tau}). 

Now let us show that all the transformed symbol matrices of the operators in Eq.\  (\ref{eq_a_state}),  namely, the $\Lambda_i$, are diagonal to first order. Let us write $\hat{\Lambda} [ \hat{D} ]$ to denote the operator $\hat{U}^\dagger \hat{D} \hat{U}$, and write $\Lambda [ \hat{D} ]$ for its Weyl symbol. First, consider the operators $\hat{I}_r$, $r = 1,3, 4, 5, 6$, which are proportional to the identity matrix. Using the operator identity

\begin{equation}
[\hat{\Lambda}(\hat{I}_r)]_{\mu\nu} = \hat{U}^\dagger_{\alpha \mu} ( \hat{I}_r  \delta_{\alpha \beta} ) \hat{U}_{\beta \nu} = \hat{I}_r \delta_{\mu \nu} - \hat{U}^\dagger_{\alpha \mu} [ \hat{U}_{\alpha \nu} , \, \hat{I}_r] \, ,
\end{equation}
we find

\begin{equation}
\label{eq_symbol_trick}
[\Lambda(\hat{I}_r)]_{\mu\nu}  = (I_r - 1/2) \delta_{\mu \nu} - i \hbar  U_{0\alpha \mu}^* \, \{ U_{0 \alpha \nu} , \, I_r \} \, , 
\end{equation}
where we have used the fact that the symbol of a commutator is a Poisson bracket. Since $U_{\alpha \mu} = \tau^{(\mu)}_\alpha$ is a function only of ${\bf J}_{12}$, and since the Poisson brackets $\{ {\bf J}_{1}, I_r \} = 0$ vanish for all $r = 1,3, 4, 5, 6$, the second term in Eq.\  (\ref{eq_symbol_trick}) vanishes. We have

\begin{equation}
[\Lambda(\hat{I}_r)]_{\mu\nu} = (I_r - 1/2) \, \delta_{\mu\nu} \, .
\end{equation}
Similarly, because $\{ {\bf J}_{1}, J_{34}^2 \} = 0$ and $\{ {\bf J}_{1}, J_{346}^2 \} = 0$, we find

\begin{equation}
[\Lambda(\hat{J}_{34}^2 ) ]_{\mu\nu} = J_{34}^2 \, \delta_{\mu\nu} \, ,  \quad \quad
[\Lambda(\hat{J}_{346}^2 ) ]_{\mu\nu} = J_{346}^2 \, \delta_{\mu\nu} \, .
\end{equation}
Now we find the symbol matrices $\Lambda({\bf \hat{J}}_{12})$ for the vector of operators ${\bf \hat{J}}_{12}$, where 

\begin{equation}
[\hat{\Lambda}({\bf \hat{J}}_{12})]_{\mu \nu} = \hat{U}^\dagger_{\alpha \mu} ( {\bf \hat{J}}_{1} \delta_{\alpha \beta}) \hat{U}_{\beta \nu} + \hbar \, \hat{U}^\dagger_{\alpha \mu} {\bf S}_{\alpha \beta} \hat{U}_{\beta \nu} \, . 
\end{equation}
After converting the above operator equation to Weyl symbols, we find 

\begin{eqnarray}
\label{eq_J_12_Lambda}
&& [\Lambda({\bf \hat{J}}_{12})]_{\mu\nu}    \\  \nonumber
&=& {\bf J}_{1} \delta_{\mu \nu} - i \hbar U_{\alpha \mu}^* \{ U_{\alpha\mu } , \, {\bf J}_{1} \}  + \hbar \, U_{\alpha \mu}^* {\bf S}_{\alpha \beta}  U_{\beta \nu} \\  \nonumber
&=& {\bf J}_{1} \delta_{\mu \nu} - i \hbar \tau^{(\mu)*}_\alpha  \{ \tau^{(\nu)}_\alpha , \, {\bf J}_{1} \}  + \hbar \, \tau^{(\mu)*}_\alpha {\bf S}_{\alpha \beta}  \tau^{(\nu)}_\beta \, . 
\end{eqnarray}
Let us denote the second term above by $T^i_{\mu \nu}$, and use (\ref{eq_tau_derivative}), the orthogonality of $\tau$, 
\begin{equation}
\tau^{(\mu)*}_\alpha  \, \tau^{(\nu)}_\alpha = \delta_{\mu \nu} \, ,
\end{equation}
to get

\begin{eqnarray}
\label{eq_T_i}
T^i_{\mu \nu} &=&  - i \hbar \tau^{(\mu)*}_\alpha \{ \tau^{(\nu)}_\alpha  , \, J_{1i} \}  \\  \nonumber
&=&  - i  \hbar \tau^{(\mu)*}_\alpha   \epsilon_{kji}   \left(   J_{1k} \frac{\partial \tau^{(\nu)}_\alpha}{ \partial J_{1j} }  \right)  \\  \nonumber
&=& \hbar ({\bf A}_{1}^{(\mu)}  \times {\bf J}_{1})_i \, \delta_{\mu \nu} + \hbar \frac{\mu {J}_{1i}}{J_{1}} \delta_{\mu \nu} - \hbar \, \tau^{(\mu)*}_\alpha {S}_{\alpha \beta}  \tau^{(\nu)}_\beta \, ,
\end{eqnarray}
where in the second equality, we have used the reduced Lie-Poisson bracket (Eq.\ (30) in \cite{littlejohn2007}) to evaluate the Poisson bracket $\{ \tau, {\bf J}_1 \}$, and in the third equality, we have used Eq.\  (\ref{eq_tau_derivative}) for $\partial \tau / \partial {\bf J}_{1}$. Notice the term involving ${\bf S}$ in $T^i_{\mu\nu}$ in Eq.\  (\ref{eq_T_i}) cancels out the same term in $\Lambda({\bf \hat{J}}_{12})$ in Eq.\  (\ref{eq_J_12_Lambda}), leaving us with a diagonal symbol matrix 

\begin{equation}
\label{eq_J12_vector}
[\Lambda({\bf \hat{J}}_{12})]_{\mu\nu} = {\bf J}_{1} \left[ 1 + \frac{\mu \hbar}{J_{1}} \right] + \hbar \, {\bf A}_{1}^{(\mu)} \times {\bf J}_{1} \, . 
\end{equation}
Taking the square, we obtain 

\begin{equation}
[ \Lambda({\bf \hat{J}}_{12}^2) ]_{\mu\nu}
=  ( J_{1}  + \mu \hbar)^2  \delta_{\mu \nu} \, . 
\end{equation}
Finally, let us look at the last three remaining operators ${\bf \hat{J}}_{\text{tot}}$ in Eq.\  (\ref{eq_Jtot_vector}). Since each of the the symbols ${\bf J}_r$ for $r = 3,4,5,6$ defined in Eq.\ (\ref{symbol_I_J}) Poisson commutes with ${\bf J}_{1}$, that is, $\{ {\bf J}_{1}, {\bf J}_r \} = 0$, we find $\Lambda({\bf \hat{J}}_r) = {\bf J}_r - i \hbar U_0^\dagger \{ U_0({\bf J}_{1}), {\bf J}_r \} = {\bf J}_r$, for $r = 3, 4, 5, 6$. Using $\Lambda({\bf \hat{J}}_{12})$ from Eq.\  (\ref{eq_J12_vector}), we obtain 
 
\begin{eqnarray}
&& [\Lambda({\bf \hat{J}}_{\text{tot}})]_{\mu\nu}     \\   \nonumber
&=& \left[  {\bf J}_{1}  \left( 1 + \frac{\mu \hbar}{J_{1}} \right) + \hbar \, {\bf A}_{1}^{(\mu)} \times {\bf J}_{1} + ({\bf J}_3 + {\bf J}_4 + {\bf J}_5 + {\bf J}_6 ) \right]  \delta_{\mu \nu} \, .
\end{eqnarray}
Therefore all $\Lambda_i$, $i = 1,\dots, 12$ are diagonal. 

Not counting the trivial eigenvalue equation for $S^2$, we have $11$ Hamilton-Jacobi equations associated with the $\Lambda_i$ for each polarization $\mu$ in the $20$ dimensional phase space  ${\mathbb C}^{10}$. It turns out that not all of them are functionally independent. In particular, the Hamilton-Jacobi equations $\Lambda(\hat{J}_{1}^2) = J_{1}^2 \hbar = ( j_{1} + 1/2 ) \hbar$ and $\Lambda(\hat{J}_{12}^2) = (J_{1} + \mu \hbar)^2 = ( j_{12} + 1/2 )^2 \hbar^2$ are functionally dependent. For them to be consistent, we must pick out the polarization $\mu = j_{12} - j_{1}$. This reduces the number of independent Hamilton-Jacobi equations for $S(x)$ from $11$ to $10$, half of the dimension of the phase space ${\mathbb C}^{10}$. These ten equations define the Lagrangian manifold associated with the action $S(x)$. 

Now let us restore the index $a$. We express the multicomponent wave function $\psi^a_\alpha(x)$ in the form of Eq.\  (\ref{eq_general_wave function}), 

\begin{equation}
\label{eq_general_wave function_a}
\psi^a_\alpha (x) = B_a(x) \, e^{i S_a(x) / \hbar} \,  \tau^a_\alpha(x, p)  \, . 
\end{equation}
Here the action $S_a(x)$ is the solution to the ten Hamilton-Jacobi equations associated with the $\mu^{\text{th}}$ entries $\lambda_i^a$ of 10 of the symbol matrices $\Lambda_i^a$, given by
\begin{eqnarray}
\label{HJ_S_a}
I_1 &=& (j_1 + 1/2) \hbar \, ,  \\    \nonumber
I_3 &=& (j_3 + 1/2) \hbar \, ,  \\  \nonumber
I_4 &=& (j_4 + 1/2) \hbar \, ,  \\  \nonumber
I_5 &=& (j_5 + 1/2) \hbar \, ,  \\  \nonumber
I_6 &=& (j_6 + 1/2) \hbar \, ,  \\  \nonumber
J_{34}^2 &=& (j_{34} + 1/2)^2 \hbar^2 \, ,  \\  \nonumber
J_{346}^2 &=& (j_{346} + 1/2)^2 \hbar^2 \, ,  \\  \nonumber
{\bf J}_{\text{tot}}^{(a)} &=& {\bf J}_{1} \left[ 1 + \frac{\mu \hbar}{J_{1}} \right] + \hbar \, {\bf A}_{1} \times {\bf J}_{1} + ({\bf J}_3 + {\bf J}_4 + {\bf J}_5 + {\bf J}_6 ) = {\bf 0} \, ,
\end{eqnarray}
and $\tau^a = \tau^{(\mu)}$ with $\mu = j_{12}-j_{1}$. Note that all the Hamiltonians in Eq.\ (\ref{HJ_S_a}) except the last three, ${\bf J}_{\text{tot}}^{(a)}$, preserve the vector value of ${\bf J}_{1}$ and ${\bf J}_{5}$ along their Hamiltonian flows.

We carry out an analogous analysis for $\psi^b(x)$. The result is
\begin{equation}
\label{eq_general_wave function_b}
\psi^b_\alpha (x) = B_b(x) \, e^{i S_b(x) / \hbar} \,  \tau^b_\alpha(x, p) \, ,
\end{equation}
where $S_b(x)$ is the solution to the following $10$ Hamilton-Jacobi equations: 

\begin{eqnarray}
I_1 &=& (j_1 + 1/2) \hbar \, ,  \\    \nonumber
I_3 &=& (j_3 + 1/2) \hbar \, ,  \\  \nonumber
I_4 &=& (j_4 + 1/2) \hbar \, ,  \\  \nonumber
I_5 &=& (j_5 + 1/2) \hbar \, ,  \\  \nonumber
I_6 &=& (j_6 + 1/2) \hbar \, ,  \\  \nonumber
J_{13}^2 &=& (j_{13} + 1/2)^2 \hbar^2 \, ,  \\  \nonumber
J_{135}^2 &=& (j_{135} + 1/2)^2 \hbar^2 \, ,  \\  \nonumber
{\bf J}_{\text{tot}}^{(b)} &=& {\bf J}_{4} \left[ 1 + \frac{\nu \hbar}{J_{4}} \right] + \hbar \, {\bf A}_{4} \times {\bf J}_{4} + ({\bf J}_1 + {\bf J}_3 + {\bf J}_5 + {\bf J}_6 ) = {\bf 0} \, .
\end{eqnarray}
Here the spinor $\tau^b = \tau_b^{(\nu)}$ satisfies 

\begin{equation}
\label{eq_JS_eigenvector_b}
({\bf J}_{4} \cdot {\bf S})_{\alpha \beta} \, ( \tau^{(\nu)}_b)_\beta = \nu J_{4} \, ( \tau^{(\nu)}_b)_\beta \, ,
\end{equation}
where $\nu = j_{24}-j_{4}$.

The vector potential ${\bf A}_{4}$ is defined by

\begin{equation}
{\bf A}_{4} = i ( \tau^b )^\dagger \, \frac{\partial \tau^b }{\partial {\bf J}_{4}} \, .
\end{equation}
Again, note that all the Hamiltonians except the last three, ${\bf J}_{\text{tot}}^{(b)}$, preserve the value of ${\bf J}_{4}$ and ${\bf J}_{6}$ along their Hamiltonian flows.

We follow the procedure described by the analysis preceding Eq.\ (69) in \cite{yu2011a} to transform the wave functions into their gauge-invariant form. The result is a gauge-invariant representation of the wave function,  

\begin{equation}
\label{wave_fctn_factorization_a}
\psi^a(x) = B_a(x) \, e^{i S_a^{6js}(x) / \hbar} \, \left[  U_a(x)  \, \tau^a(x_0) \right]   \, . 
\end{equation}
where the action $S_a^{6js}(x)$ is the integral of $p \, dx$ starting at a point $z_0$, which is the lift of a reference point $x_0$ in the Lagrangian manifold ${\mathcal L}_a^{6js}$. The Lagrangian manifold ${\mathcal L}_a^{6js}$ is defined by the following equations:

\begin{eqnarray}
\label{HJ_S_a_6js}
I_1 &=& (j_1 + 1/2) \hbar \, ,  \\    \nonumber
I_3 &=& (j_3 + 1/2) \hbar \, ,  \\  \nonumber
I_4 &=& (j_4 + 1/2) \hbar \, ,  \\  \nonumber
I_5 &=& (j_5 + 1/2) \hbar \, ,  \\  \nonumber
I_6 &=& (j_6 + 1/2) \hbar \, ,  \\  \nonumber
J_{34}^2 &=& (j_{34} + 1/2)^2 \hbar^2 \, ,  \\  \nonumber
J_{346}^2 &=& (j_{346} + 1/2)^2 \hbar^2 \, ,  \\  \nonumber
{\bf J}_{\text{tot}} &=& {\bf J}_1  + {\bf J}_3 + {\bf J}_4 + {\bf J}_5 + {\bf J}_6  = {\bf 0} \, .
\end{eqnarray}
The rotation matrix $U_a(x)$ that appears in Eq.\ (\ref{wave_fctn_factorization_a}) is determined by the $\SO(3)$ rotation that transforms the shape configuration of ${\bf J}_{1}$ and ${\bf J}_5$ at the reference point $z_0 = (x_0, p(x_0))$ on ${\mathcal L}_a^{6js}$ to the shape configuration of ${\bf J}_{1}$ and ${\bf J}_5$ at the point $z = (x, p(x))$ on ${\mathcal L}_a^{6js}$. Here ${\bf J}_{1}$ and ${\bf J}_5$ are functions of $z$ and are defined in Eq.\ (\ref{symbol_I_J}).

Similarly, the multicomponent wave function for the state $\ket{b}$ has the following form,

\begin{equation}
\label{wave_fctn_factorization_b}
\psi^b(x) = B_b(x) \, e^{i S_b^{6js}(x) / \hbar} \, \left[ U_b(x)   \, \tau^b(x_0)  \right] \, , 
\end{equation}
where the action $S_b^{6js}(x)$ is the integral of $p \, dx$ starting at a point that is the lift of $x_0$ onto the Lagrangian manifold ${\mathcal L}_b^{6js}$. The Lagrangian manifold ${\mathcal L}_b^{6js}$ is defined by the following equations:

\begin{eqnarray}
\label{HJ_S_b_6js}
I_1 &=& (j_1 + 1/2) \hbar \, ,  \\    \nonumber
I_3 &=& (j_3 + 1/2) \hbar \, ,  \\  \nonumber
I_4 &=& (j_4 + 1/2) \hbar \, ,  \\  \nonumber
I_5 &=& (j_5 + 1/2) \hbar \, ,  \\  \nonumber
I_6 &=& (j_6 + 1/2) \hbar \, ,  \\  \nonumber
J_{13}^2 &=& (j_{13} + 1/2)^2 \hbar^2 \, ,  \\  \nonumber
J_{135}^2 &=& (j_{135} + 1/2)^2 \hbar^2 \, ,  \\  \nonumber
{\bf J}_{\text{tot}} &=& {\bf J}_1 + {\bf J}_3 + {\bf J}_4 + {\bf J}_5 + {\bf J}_6  = {\bf 0} \, .
\end{eqnarray}
The rotation matrix $U_b(x)$ that appears in Eq.\ (\ref{wave_fctn_factorization_b}) is determined by the $\SO(3)$ rotation that transform the shape configuration of ${\bf J}_{4}$ and ${\bf J}_6$ at the reference point $z_0 = (x_0, p(x_0))$ on ${\mathcal L}_b^{6js}$ to the shape configuration of ${\bf J}_{4}$ and ${\bf J}_6$ at the point $z = (x, p(x))$ on ${\mathcal L}_b^{6js}$.

Taking the inner product of the wave functions, and treating the spinors as part of the slowly varying amplitudes, we find 

\begin{eqnarray}
 \braket{b|a}  
&=&  e^{i\kappa} \sum_k    \Omega_k  \, \text{exp} \{i [S_a^{6js}(z_k) - S_b^{6js}(z_k) - \mu_k \pi /2 ] / \hbar \}   \nonumber  \\
&&  \left( U_b^{0k} \tau^b(z_0)\right)^\dagger  \left(U_a^{0k} \tau^a(z_0)\right)  .
\label{eq_general_formula}
\end{eqnarray}
In the above formula, the sum is over the components  of the intersection set ${\mathcal M}_k$ between the two Lagrangian manifolds ${\mathcal L}_a^{6js}$ and ${\mathcal L}_b^{6js}$. The point $z_k$ is any point in the $k$th component. The amplitude $\Omega_k$ and the Maslov index $\mu_k$ are the results of doing the stationary phase approximation of the inner product without the spinors. Each rotation matrix $U_a^{0k}$ is determined by a path $\gamma^{a (0k)}$ that goes from $z_0$ to $z_k$ along ${\mathcal L}_a^{6js}$, and $U_b^{0k}$ is similarly defined. The formula  (\ref{eq_general_formula}) is independent of the choice of $z_k$, because any other choice $z_k'$ will multiply both $U_a^{0j}$ and $U_b^{0j}$ by the same additional rotation matrix which cancels out in the product $(U_b^{0k})^\dagger U_a^{0k}$.

\section{\label{sec_lagrangian_manifolds}The Lagrangian Manifolds}

\begin{figure}[tbhp]
\begin{center}
\includegraphics[width=0.50\textwidth]{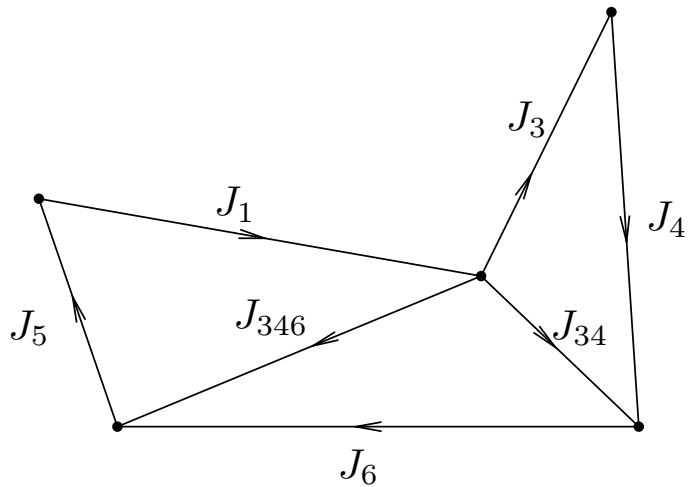}
\caption{The configuration of a point on ${\mathcal L}_a^{6js}$, projected onto the angular momentum space $\Lambda_{5j}$, and viewed in a single copy of ${\mathbb R}^3$.}
\label{fig_9j_config_a}
\end{center}
\end{figure}

We now analyze the Lagrangian manifolds   ${\mathcal L}_a^{6js}$ and ${\mathcal L}_b^{6js}$, defined by the Hamilton-Jacobi equations Eq.\  (\ref{HJ_S_a_6js})  and  Eq.\  (\ref{HJ_S_b_6js}), respectively. We focus on  ${\mathcal L}_a^{6js}$ first, since the treatment for  ${\mathcal L}_b^{6js}$ is analogous. Let $\pi: \Phi_{5j} \rightarrow \Lambda_{5j}$ denote the projection of the large phase space  $\Phi_{5j} = ({\mathbb C}^2)^5$ onto the angular momentum space $\Lambda_{5j} = ({\mathbb R}^3)^5$, through the functions ${\bf J}_{ri}$, $r = 1,3,4,5, 6$.  The first six equations, $I_r = j_r + 1/2$, $r = 1,3,4,5,6$ fix the lengths of the five vectors $|{\bf J}_r| = J_r$, $r = 1,3,4,5, 6$. The three equations for the total angular momentum, 

\begin{equation}
\label{eq_total_J_0}
{\bf J}_{\text{tot}} = {\bf J}_1  + {\bf J}_3 + {\bf J}_4 + {\bf J}_5 + {\bf J}_6  = {\bf 0} \, ,
\end{equation}
constrains the five vectors ${\bf J}_i$, $i = 1, \dots, 6$ to form a close polygon. The remaining two equations 

\begin{eqnarray}
J_{34}^2 &=& (j_{34} + 1/2)^2 \hbar^2 \, ,  \\ 
J_{346}^2 &=& J_{15}^2 = (j_{346} + 1/2)^2 \hbar^2 \, ,  
\end{eqnarray}
put the vectors ${\bf J}_3, {\bf J}_4$ into a 3-4-34 triangle, and put the vectors ${\bf J}_1, {\bf J}_5$ into a 1-5-346 triangle. Thus, the vectors form a butterfly shape, illustrated in Fig.\ \ref{fig_9j_config_a}. This shape has two wings $(J_3,  J_4, J_{34})$ and $(J_1, J_5, J_{346})$ that are free to rotate about the $J_{34}$ and $J_{346}$ edges, respectively. Moreover, the Hamilton-Jacobi equations are also invariant under an overall rotation of the vectors. Thus the projection of ${\mathcal L}_a^{6js}$ onto the angular momentum space is diffeomorphic to $\U(1)^2 \times \Ortho(3)$.

The orbit of the group $\U(1)^5$ generated by $I_r$, $r = 1,3,4,5, 6$ is a $5$-torus. Thus ${\mathcal L}_a^{6js}$ is a $5$-torus bundle over a sub-manifold described by the butterfly configuration in Fig.\ \ref{fig_9j_config_a}. Altogether there is a $\U(1)^7 \times \SU(2)$ action on ${\mathcal L}_a^{6js}$. If we denote coordinates on $\U(1)^7 \times \SU(2)$ by $(\psi_1, \psi_2, \psi_3, \psi_4, \psi_6, \theta_{34}, \theta_{346}, u)$, where $u \in \SU(2)$ and where the five angles are the $4\pi$-periodic evolution variables corresponding to $(I_1, I_3, I_4, I_5, I_6, {\bf J}_{34}^2, {\bf J}_{346}^2)$, respectively, then the isotropy subgroup is generated by three elements, say $x=(2 \pi, 2 \pi, 2 \pi, 2 \pi, 2 \pi, 0, 0, -1)$,  $y = (2 \pi, 0, 0, 2 \pi, 2 \pi, 2 \pi, 0, -1)$, and $z = ( 2 \pi, 0, 0, 2 \pi, 0, 0, 2 \pi, -1)$. The isotropy subgroup itself is an Abelian group of eight elements, $({\mathbb Z}_2)^3 = \{ e, x, y, z, xy, xz, yz, xyz \}$. Thus the manifold ${\mathcal L}_a^{6js}$  is topologically $\U(1)^7 \times \SU(2) / ({\mathbb Z}_2)^3$. The analysis for ${\mathcal L}_b^{6js}$ is the same. 

Now it is easy to find the invariant measure on ${\mathcal L}_a^{6js}$ and ${\mathcal L}_b^{6js}$. It is $d \psi_1 \wedge d \psi_3 \wedge d\psi_4 \wedge d\psi_5 \wedge d \psi_6 \wedge d\theta_{34} \wedge d \theta_{346} \wedge du$, where $du$ is the Haar measure on $\SU(2)$. The volumes $V_A$ of ${\mathcal L}_a^{6js}$ and $V_B$ of ${\mathcal L}_b^{6js}$ with respect to this measure are 

\begin{equation}
V_A = V_B = \frac{1}{8} \, (4 \pi)^7 \times 16 \pi^2 = 2^{15} \pi^9  \, , 
\end{equation}
where the $1/8$ factor compensates for the $8$-element isotropy subgroup.

\begin{figure}[tbhp]
\begin{center}
\includegraphics[width=0.48\textwidth]{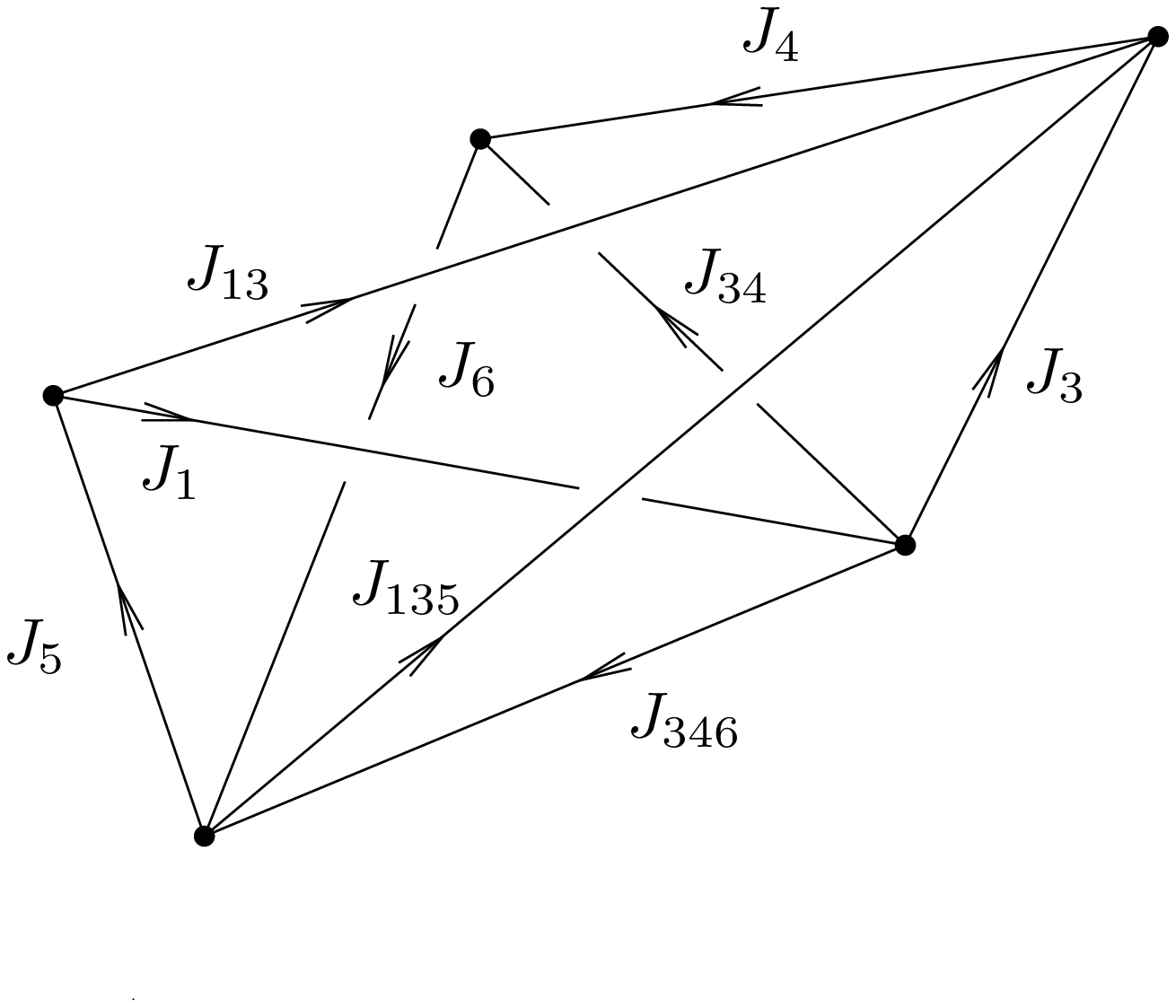}
\caption{The configuration of a point on the intersection set $I_{21}$, projected onto the angular momentum space $\Lambda_{5j}$.}
\label{fig_6js_config_I_21}
\end{center}
\end{figure}

We now examine the intersections of ${\mathcal L}_a^{6js}$ and ${\mathcal L}_b^{6js}$ in detail. Because the two lists of Hamilton-Jacobi equations  Eq.\  (\ref{HJ_S_a_6js})  and  Eq.\  (\ref{HJ_S_b_6js}) share the common equations $I_r = j_r + 1/2$, $r = 1,2,3,4,6$, the intersection in the large phase space $\Phi_{5j}$ is a $5$-torus fiber bundle over the intersection of the projections in the angular momentum space $\Lambda_{5j}$. The intersections of the projections in $\Lambda_{5j}$ require the five vectors ${\bf J}_r$, $r = 1,3,4,5,6$, to satisfy

\begin{eqnarray}
| {\bf J}_r | = J_r \, , \quad\quad\quad  \sum_r  {\bf J}_r = {\bf 0}  \, ,    \nonumber  \\ 
\label{eq_vector_equations}
|{\bf J}_3 + {\bf J}_4 | = J_{34}  \, , \quad \quad  |{\bf J}_3 + {\bf J}_4 + {\bf J}_6 | = J_{346}  \, ,   \\  \nonumber
|{\bf J}_1 + {\bf J}_3 | = J_{13}  \, , \quad \quad  |{\bf J}_1 + {\bf J}_3 + {\bf J}_5 | = J_{135}   \, .
\end{eqnarray}
These conditions imply that the six edges  $J_3, J_{135}, J_{346}, J_4, J_{34}, J_6$ form a tetrahedron, and the six edges $J_3, J_{135}, J_{346}, J_1, J_{13}, J_5$ form another tetrahedron. The two tetrahedra share the common face $(J_3, J_{135}, J_{346})$, as illustrated in Fig.\ \ref{fig_6js_config_I_11}. We can use the procedure explained in the appendix of \cite{littlejohn2009} to construct a tetrahedron with the six edge lengths $J_3, J_{135}, J_{346}, J_4, J_{34}, J_6$. This procedure gives us the vectors ${\bf J}_3, {\bf J}_{135}, {\bf J}_{346}, {\bf J}_1, {\bf J}_5, {\bf J}_{13}$. Then we use the following three conditions

\begin{eqnarray}
{\bf J}_4 \cdot {\bf J}_4 &=& J_4^2  \nonumber  \\
\label{eq_J4_conditions}
{\bf J}_4 \cdot {\bf J}_3 &=& \frac{1}{2} ( J_{34}^2 - J_4^2 - J_3^2 )  \\  \nonumber
{\bf J}_4 \cdot {\bf J}_{135} &=& \frac{1}{2} ( J_4^2 + J_{135}^2 - J_6^2 ) 
\end{eqnarray}
to solve for the vector ${\bf J}_4$. In general, there are two solutions for ${\bf J}_4$, corresponding to the two orientations of the second tetrahedron. See Fig.\ \ref{fig_6js_config_I_11} and Fig.\ \ref{fig_6js_config_I_21}. Once we have ${\bf J}_4$, we then use

\begin{eqnarray}
{\bf J}_{34} &=& {\bf J}_3 + {\bf J}_4  \\
{\bf J}_{6} &=& {\bf J}_{346} - {\bf J}_{34} 
\end{eqnarray}
to construct the second tetrahedron with the six edges $J_3, J_{135}, J_{346}, J_4, J_{34}, J_6$.

The two solutions to Eq.\ (\ref{eq_J4_conditions}) give rise two vector configurations that are not related by an $\Ortho(3)$ symmetry. This means that there are four vector configurations satisfying Eqs. (\ref{eq_vector_equations}) that are not related by an $\SO(3)$ symmetry. The intersections in $\Phi_{5j}$ are the lifts of the intersections in $\Lambda_{5j}$. Therefore, the intersection of ${\mathcal L}_a^{6js}$  consists of four disconnected subsets, where each subset is a $5$-torus bundle over $\SO(3)$. Let us denote the two sets corresponding to the configuration in which the two tetrahedra are on opposite sides of the $(J_3, J_{135}, J_{346})$ triangle and its mirror image by $I_{11}, I_{12}$, and denote the configuration in which the two tetrahedra are on the same side of the triangle $(J_3, J_{135}, J_{346})$ and its mirror image by $I_{21}, I_{22}$. The vector configuration for a typical point in $I_{11}$ is illustrated in Fig.\ \ref{fig_6js_config_I_11}, and the vector configuration for a typical point in $I_{21}$ is illustrated in Fig.\ \ref{fig_6js_config_I_21}. Each intersection set is an orbit of the group $\U(1)^5 \times \SU(2)$, where $\U(1)^5$ represent the phases of the five spinors and $\SU(2)$ is the diagonal action generated by ${\bf J}_{\rm tot}$.

\begin{figure}[tbhp]
\begin{center}
\includegraphics[width=0.48\textwidth]{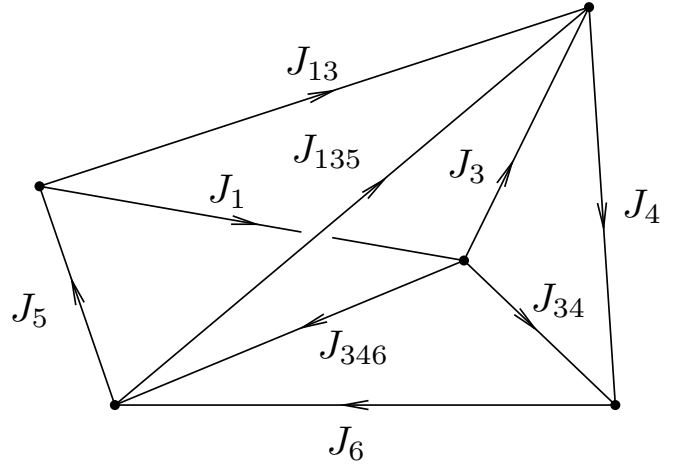}
\caption{The configuration of a point on the intersection set $I_{11}$, projected onto the angular momentum space $\Lambda_{5j}$.}
\label{fig_6js_config_I_11}
\end{center}
\end{figure}

The isotropy subgroup of this group action is ${\mathbb Z}_2$, generated by the element $ \; $ $(2 \pi, 2 \pi, 2\pi, 2\pi, 2\pi, -1)$, in coordinates $(\psi_1, \psi_3, \psi_4, \psi_5, \psi_6, u)$ for the group $\U(1)^5 \times \SU(2)$, where $u \in \SU(2)$. The volume of the intersection manifold $I_{11}$, $I_{12}$, $I_{21}$, or $I_{22}$, with respect to the measure $d \psi_1 \wedge  d\psi_3 \wedge d\psi_4 \wedge d \psi_5 \wedge d \psi_6  \wedge du$, is 

\begin{equation}
V_I = \frac{1}{2} (4 \pi)^5 \times 16 \pi^2 = 2^{13} \pi^7   \, ,
\end{equation}
where the $1/2$ factor compensates for the two element isotropy subgroup. 

The amplitude determinant is given in terms of a determinant of Poisson brackets among distrinct Hamiltonians between the two lists of Hamilton-Jacobi equations in Eq.\  (\ref{HJ_S_a_6js})  and  Eq.\  (\ref{HJ_S_b_6js}). In this case, those are $(J_{34}, J_{346})$ from Eq.\  (\ref{HJ_S_a_6js})  and $(J_{13}, J_{135})$ from  Eq.\  (\ref{HJ_S_b_6js}). Thus the determinant of Poisson brackets is
\begin{eqnarray}
&& \left|
  \begin{array}{cc}
    \{J_{34}, \, J_{13} \} & \{J_{34}, \, J_{135} \}   \\ 
    \{J_{346}, \, J_{13} \}  & \{J_{346}, \, J_{135} \}   \\ 
  \end{array} 
  \right|  \nonumber  \\   \nonumber
&=& \frac{1}{J_{34} J_{346} J_{13} J_{135} } \,
  \left|
  \begin{array}{cc}
    V_{341} & V_{346}  \\ 
    V_{135} & V_{3(46)(15)}   \\ 
  \end{array} 
  \right|  \\
  &=& \frac{1}{J_{34} J_{346} J_{13} J_{135} }  | V_{135} V_{346}   |  \, , 
\end{eqnarray}
where, in the last equality, we have used $V_{3(15)(46)} = {\bf J}_3 \cdot ({\bf J}_{46} \times {\bf J}_{15}) = 0$, since the edges $(J_3, J_{46}, J_{15})$ form a triangle. Here $V_{ijk}$ is six times the volume of the tetrahedron generated from $J_i, J_j, J_k$, and is given by

\begin{equation}
V_{ijk} = {\bf J}_i \cdot ({\bf J}_j \times {\bf J}_k)  \, . 
\end{equation}

The amplitude $\Omega_k$ in Eq.\  (\ref{eq_general_formula}) can be inferred from Eq.\ (10) in \cite{littlejohn2010a}. In the present case, each $\Omega_k$ has the same expression $\Omega$. It is 

\begin{eqnarray}
\Omega &=& \frac{(2 \pi i ) V_I}{\sqrt{V_A V_B} }  \, \frac{\sqrt{J_{34} J_{346} J_{13} J_{135}} }{\sqrt{| V_{135} V_{346}  |} }  \nonumber   \\  \nonumber
&=& \frac{(2 \pi i ) 2^{13} \pi^7 }{ 2^{15} \pi^9 }  \, \frac{\sqrt{J_{34} J_{346} J_{13} J_{135}} }{\sqrt{| V_{135} V_{346} |} }     \\
&=& \frac{i \sqrt{J_{34} J_{346} J_{13} J_{135}}}{2 \pi \sqrt{| V_{135} V_{346}  |} }  \, . 
\label{eq_amplitude}
\end{eqnarray}

We now outline the calculation of the relative phase between the exponents $S_a(z_{12}) - S_b(z_{12})$ and $S_a(z_{11}) - S_b(z_{11})$, which can be written as an action integral 

\begin{equation}
S^{(1)} = (S_a(z_{12}) - S_b(z_{12}) ) - (S_a(z_{11}) - S_b(z_{11}) )  = \oint \, p \, dx \, 
\end{equation}
around a closed loop that goes from $z_{11}$ to $z_{12}$ along ${\mathcal L}_a^{6js}$ and then back along ${\mathcal L}_b^{6js}$.

\begin{figure}[tbhp]
\begin{center}
\includegraphics[width=0.48\textwidth]{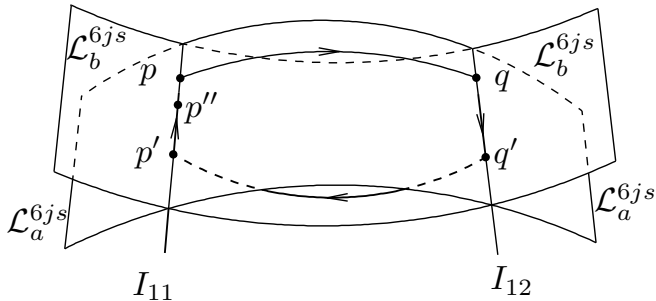}
\caption{The loop from a point $p \in I_{11}$ to $q \in I_{12}$ along ${\mathcal L}_a^{6js}$, and then to $q' \in I_{12}$ along $I_{12}$, and then to $p' \in I_{11}$ along ${\mathcal L}_b^{6js}$, and finally back to $p''$ and then to $p$ along $I_{11}$.}
\label{fig_loop_large_space}
\end{center}
\end{figure}

\begin{figure}[tbhp]
\begin{center}
\includegraphics[width=0.48\textwidth]{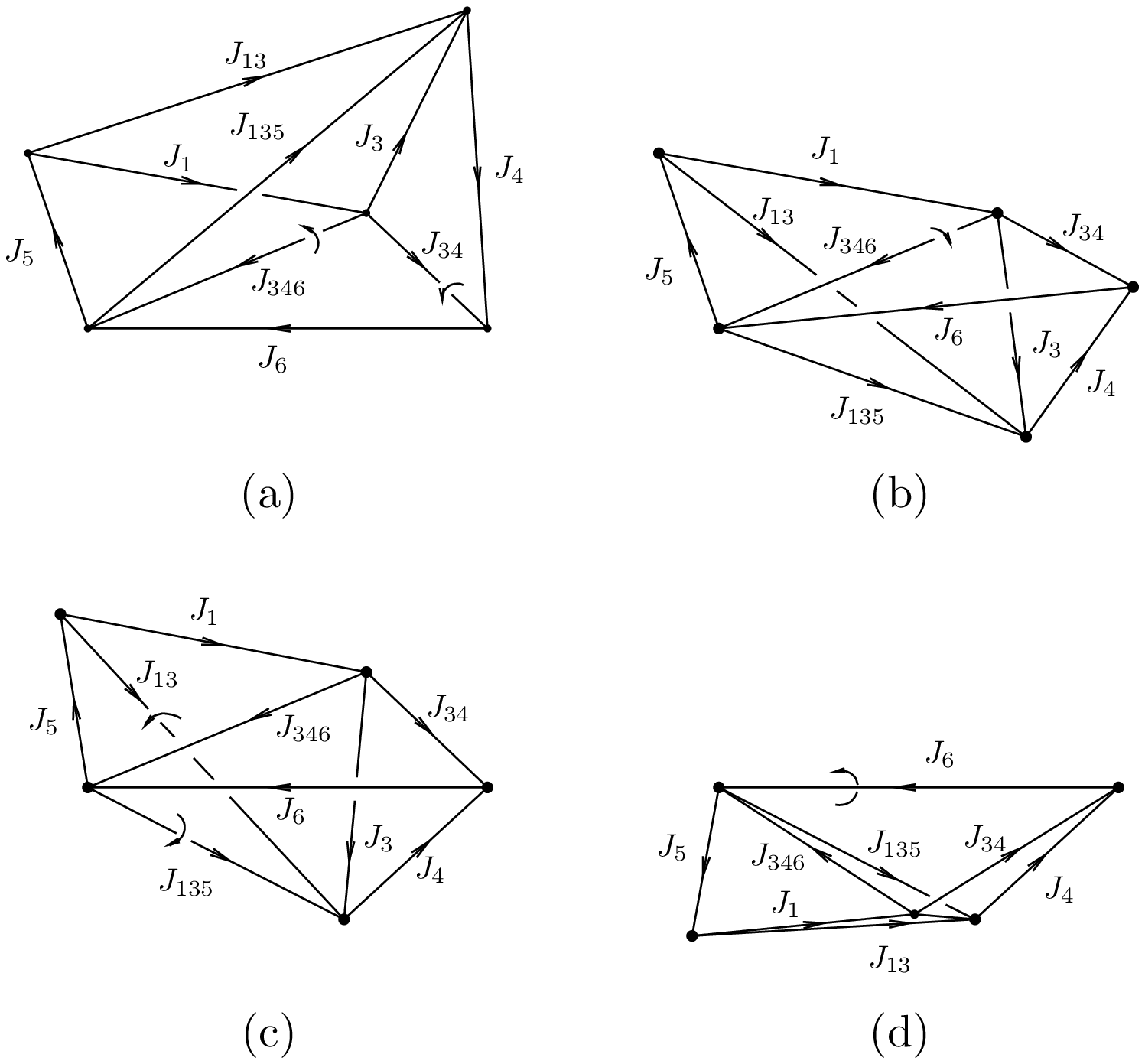}
\caption{The loop from Fig.\ \ref{fig_loop_large_space} projected onto a loop (a)$\rightarrow$(b)$\rightarrow$(c)$\rightarrow$(d)$\rightarrow$(a) in $\Lambda_{5j}$.} 
\label{fig_loop_small_space}
\end{center}
\end{figure}

We shall construct the closed loop giving the relative phase $S^{(1)}$ by following the Hamiltonian flows of various observables. This loop consists of four paths, and it is illustrated in the large phase space $\Phi_{5j}$ in Fig.\ \ref{fig_loop_large_space}. The loop projects onto a loop in the angular momentum space $\Lambda_{5j}$, which is illustrated in Fig.\ \ref{fig_loop_small_space}. We take the starting point $p \in I_{11}$ of Fig.\ \ref{fig_loop_large_space} to lie in the $5$-torus fiber above a solution of Eq.\  (\ref{eq_vector_equations}). The projection of $p$ in $\Lambda_{5j}$ is illustrated in part (a) of Fig.\ \ref{fig_loop_small_space}.

First we follow the ${\bf J}_{34}^2$-flow and then the ${\bf J}_{346}^2$-flow to trace out a path that takes us along ${\mathcal L}_a^{6js}$ from a point $p$ in $I_{11}$ to a point $q$ in $I_{12}$. Let the angles of rotations be $2 \phi_{34}$ and $2 \phi_{346}$, respectively, where $\phi_{12}$ is the angle between the triangles 3-4-34 and 34-6-346, and $\phi_{346}$ is the angle between the triangles 34-6-346 and 1-5-346. These rotations effectively reflect all five vectors ${\bf J}_r$, $r = 1,3,4,5,6$ across the triangle 1-5-346, taking us from part (a) to part (b) of Fig.\ \ref{fig_loop_small_space}.

Next, we follow the Hamiltonian flow generated by $-{\bf j}_{346} \cdot {\bf J}_{\rm tot}$ along $I_{12}$, which generates an overall rotation of all the vectors around $- {\bf j}_{346}$. Let the angle of rotation be $2 \phi_{346}$ defined above. This brings the triangle 34-6-346 back to its original position. However, the triangle 1-5-346 is now rotated to the other side of triangle 34-6-346, as illustrated in part (c) of Fig.\  \ref{fig_loop_small_space}. Effectively, the actions on all five vectors ${\bf J}_r$, $r = 1,3,4,5,6$ from part (a) to part (c) of Fig.\ \ref{fig_loop_small_space} has been to reflect them across the 34-6-346 triangle.

To go back to a point $p'$ in $I_{11}$, we follow the ${\bf J}_{13}^2$-flow and ${\bf J}_{135}^2$-flow along ${\mathcal L}_b^{6js}$. Let the angle of rotations be $2\phi_{13}$ and $2\phi_{135}$, respectively, where $\phi_{13}$ is the angle between the triangle 1-3-13 and the triangle 13-5-135, and $\phi_{135}$ is the angle between the triangle 13-5-135 and the triangle 4-6-135. These rotations effectively reflect all the vectors across the 4-6-135 triangle, taking us from part (c) to part (d) of Fig.\ \ref{fig_loop_small_space}. Thus we arrive at a point $p' \in I_{11}$. We now use the fact that the product of two reflections is a rotation about the intersection of the two reflection planes. We note that the vector ${\bf J}_6$ is stationary under the reflection across the 34-6-346 plane, as well as across the 4-6-135 plane. Thus, the final rotation that brings all the vectors back to their original positions is generated by $-{\bf j}_{6} \cdot {\bf J}_{\rm tot}$ along $I_{11}$. It is an overall rotation of all the vectors about $- {\bf j}_{6}$ by an angle $2 \phi_{6}$, where $\phi_6$ is the angle between the 34-6-346 triangle and the 4-6-135 triangle. This rotation takes us from part (d) back to part (a) of Fig.\ \ref{fig_loop_small_space}. We denote the final point by $p''$ in the large phase space. The points $p$ and $p''$ have the same projection in the angular momentum space $\Lambda_{5j}$. Thus the two points $p$ and $p''$ differ only by the phases of the five spinors, which can be restored by following the Hamiltonian flows of $(I_1, I_3, I_4,  I_5, I_6)$. This constitutes the last path from $p''$ to $p$.

To summarize the rotational history in the angular momentum space, we have applied the rotations

\begin{eqnarray}
\label{ch8: eq_rotations}
&& R( - {\bf j}_{6}, 2 \phi_{6}) R_{135} ({\bf j}_{135}', 2 \phi_{135}) R_{13}({\bf j}_{13}', 2 \phi_{13})  R( - {\bf j}_{346}, 2 \phi_{346}) \nonumber  \\  
&& R_{346} ({\bf j}_{346}, 2 \phi_{346}) R_{34}({\bf j}_{34}, 2 \phi_{34})  \, , 
\end{eqnarray}
where $R_{34}$ acts only on ${\bf J}_3$ and ${\bf J}_4$, $R_{346}$ acts only on ${\bf J}_3$, ${\bf J}_4$, and ${\bf J}_6$, $R_{13}$ acts only on ${\bf J}_1$ and ${\bf J}_3$, $R_{135}$ acts only on ${\bf J}_1$, ${\bf J}_3$, and ${\bf J}_5$, and $R( - {\bf j}_{346}, 2 \phi_{346} )$, $R( - {\bf j}_6, 2 \phi_6 )$ acts on all five vectors. The corresponding $\SU(2)$ rotations, with the same axes and angles, take us from point $p$ in Fig.\ \ref{fig_loop_large_space} to another point $p''$ along the sequence $p \rightarrow q \rightarrow q'  \rightarrow p' \rightarrow p''$.

To compute the final five phases required to close the loop, we use the Hamilton-Rodrigues formula \cite{whittaker1960}, in the same way as Eq.\ (46) in \cite{littlejohn2010b}. Let us start with vector ${\bf J}_4$. The action of the rotations on this vector can be written

\begin{eqnarray}
&& R(-{\bf j}_{6}, 2 \phi_{6}) R(- {\bf j}_{346}, 2 \phi_{346})  \nonumber  \\
&& R_{346}({\bf j}_{346}, 2 \phi_{346}) R_{34}({\bf j}_{34}, 2 \phi_{34}) {\bf J}_4  \nonumber \\
&=& R(-{\bf j}_{6}, 2 \phi_{6})  R_{34}({\bf j}_{34}, 2 \phi_{34}) {\bf J}_4  \nonumber \\
&=& R({\bf j}_{4}, 2 \phi_{4})   {\bf J}_4  \nonumber \\
&=& {\bf J}_4  \, . 
\label{eq_J4_rotation}
\end{eqnarray}
where we have used the Hamilton-Rodrigues formula in the second equality. Thus, we find that the product of the rotations acting on ${\bf J}_4$ is $R({\bf j}_4, 2 \phi_4)$, where $\phi_4$ is the angle between the triangle 3-4-34 and the triangle 4-6-135. We can lift the rotations up to $\SU(2)$ with the same axis and angle. Its action on the spinor at $p$ is a pure phase. To undo this pure phase, we follow the Hamiltonian flow of $I_4$ by an angle $-2 \phi_4$, modulo $2 \pi$.

For the vector ${\bf J}_6$, the rotations acting on it is simple. 

\begin{eqnarray}
&& R(-{\bf j}_{6}, 2 \phi_{6}) R(- {\bf j}_{346}, 2 \phi_{346}) R_{346}({\bf j}_{346}, 2 \phi_{346}) {\bf J}_6  \nonumber \\
&=& R(-{\bf j}_{6}, 2 \phi_{6}) {\bf J}_6 \nonumber \\
&=& {\bf J}_6
\end{eqnarray}
We can lift the rotations up to $\SU(2)$ with the same axis and angle. Its action on the spinor at $p$ is a pure phase. To undo this pure phase, we follow the Hamiltonian flow of $I_6$ by an angle $-2 \phi_6$.

Similarly, we can find the rotations acting on ${\bf J}_1, {\bf J}_3, {\bf J}_5$, and proceed to calculate the action integral as in \cite{littlejohn2010b}. Instead, we will take a shortcut and use the fact that the two Lagrangian manifolds  ${\mathcal L}_a^{6js}$ and ${\mathcal L}_b^{6js}$ describe the WKB wave functions associated with the product of two $6j$ symbols on the right hand side of Eq.\ (\ref{eq_jahn_A8}). The asymptotic limit of a product of two $6j$ symbol can be easily derived from the Ponzano-Regge formula for a single $6j$ symbol. 

\begin{eqnarray}
\label{eq_double_PR_formula}
&& \left\{
   \begin{array}{ccc}
    j_{346} & j_3 & j_{135}  \\ 
    j_4 & j_6 & j_{34}   \\ 
  \end{array} 
  \right \}  \, 
\left\{
   \begin{array}{ccc}
    j_{346} & j_3 & j_{135}  \\ 
    j_{13}& j_5 & j_{1}   \\ 
  \end{array} 
  \right \}    \\   \nonumber
 &=&  \frac{1}{2 \pi  \sqrt{| V_{135} V_{346} |}}  \cos \left(S_2 + \frac{\pi}{4} \right)   \cos \left(S_1 + \frac{\pi}{4} \right)    \\  \nonumber
 &=& \frac{1}{4 \pi  \sqrt{| V_{135} V_{346} |}} \left[ \cos \left(  S_1 + S_2 + \frac{\pi}{2} \right)  + \cos \left(  S_1 - S_2  \right)  \right]
\end{eqnarray}
where the Ponzano-Regge phases $S_1$ and $S_2$ are 

\begin{equation}
\label{eq_S_PR_346}
S_1 = J_4 \psi_4 + J_6 \psi_6 + J_{34} \psi_{34} + J_3 \psi_3^{(1)} + J_{135} \psi_{135}^{(1)} + J_{346} \psi_{346}^{(1)} 
\end{equation}

\begin{equation}
\label{eq_S_PR_135}
S_2 = J_1 \psi_1 + J_5 \psi_5 + J_{15} \psi_{15} + J_3 \psi_3^{(2)} + J_{135} \psi_{135}^{(2)} + J_{346} \psi_{346}^{(2)} 
\end{equation}

Here $\psi_4, \psi_6 , \psi_{34} , \psi_3^{(1)}, \psi_{135}^{(1)}, \psi_{346}^{(1)}$ are the exterior dihedral angles of the tetrahedron formed by the six edges $J_4, J_6, J_{34}, J_3, J_{135}, J_{346}$, and $\psi_1, \psi_5, \psi_{15}, \psi_3^{(2)}, \psi_{135}^{(2)}, \psi_{346}^{(2)} $ are exterior dihedral angles of the tetrahedron formed by the six edges $J_1, J_5, J_{15}, J_3, J_{135}, J_{346}$.  These two tetrahedra are illustrated in Fig.\ \ref{fig_6js_config_I_11}.

The action integral $\int p \, dx$ along the loop $p \rightarrow q \rightarrow q'  \rightarrow p' \rightarrow p'' \rightarrow p$ in Fig.\ \ref{fig_loop_large_space}  can then be read off from Eq.\ (\ref{eq_double_PR_formula}). It is given by 

\begin{equation}
\label{eq_action_integral}
S^{(1)} = 2 ( S_1 + S_2 )  \, .
\end{equation}
The action integral along a similar closed loop from $I_{21}$ to $I_{22}$ and back to $I_{21}$ is 

\begin{equation}
\label{eq_action_integral_2}
S^{(2)} =  2 ( S_1 - S_2 )  \, .
\end{equation}
The Maslov indices $\mu_1 = -2$ and $\mu_2 = 0$ can also be read off from Eq.\ (\ref{eq_double_PR_formula}). Putting the amplitudes $\Omega$ from Eq.\  (\ref{eq_amplitude}) and the relative actions $S^{(1)}$  and $S^{(2)}$ and Maslov indices $\mu_1$ and $\mu_2$ into Eq.\  (\ref{eq_general_formula}), we find 

\begin{widetext}
\begin{eqnarray}
 \braket{b|a}  
&=&  \frac{\sqrt{J_{34} J_{346} J_{13} J_{135}}}{2 \pi \sqrt{| V_{135} V_{346}  |} }  \{  e^{i\kappa_1}  \left[ ( \tau^b(z_{11}) )^\dagger (\tau^a(z_{11}))  +  e^{i ( S^{(1)} + \pi ) / \hbar}   \left( U_b^{(1)} \tau^b(z_{11}) \right)^\dagger  \left(U_a^{(1)} \tau^a(z_{11})\right)     \right]    
\label{eq_general_formula_2}  \nonumber  \\   
&& +  e^{i\kappa_2}  \left[ ( \tau^b(z_{21}) )^\dagger (\tau^a(z_{21}))  +  e^{i ( S^{(2)}) / \hbar}   \left( U_b^{(2)} \tau^b(z_{21}) \right)^\dagger  \left(U_a^{(2)} \tau^a(z_{21})\right)     \right]  \}  
\label{eq_general_formula_1}
\end{eqnarray}
\end{widetext}
Here we have factored out two arbitrary phases $e^{i \kappa_1}$ and $e^{i \kappa_2}$ for the two pairs of stationary phase contributions. The rotation matrices $U_a^{(i)}$, $i = 1,2$ are determined by the paths from $z_{i1}$ to $z_{i2}$ along ${\mathcal L}_a^{6js}$. Similarly the rotation matrices $U_b^{(i)}$, $i = 1,2$, are determined by the paths from $z_{i1}$ to $z_{i2}$ along ${\mathcal L}_b^{6js}$.

\section{\label{sec_spinor_product}The Spinor Products}

\begin{figure}[tbhp]
\begin{center}
\includegraphics[width=0.30\textwidth]{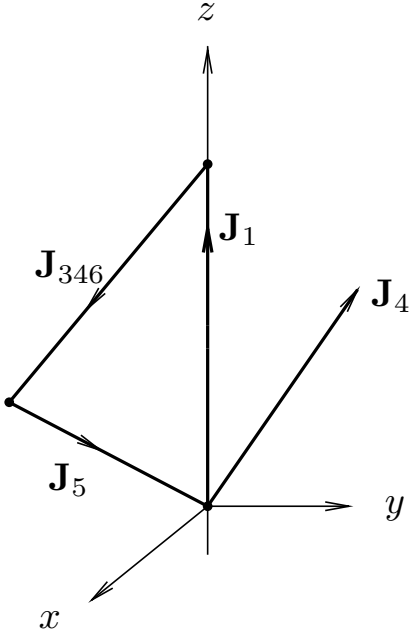}
\caption{The vector configuration at the point $z_{11}$ in $I_{11}$.}
\label{fig_config_z11}
\end{center}
\end{figure}

We now calculate the spinor products in Eq.\ (\ref{eq_general_formula_1}). We choose the vector configurations associated with $z_{11}$ to correspond to a particular orientation of the vectors. We put ${\bf J}_{1}$ along the $z$-axis, and put ${\bf J}_5$ inside the $xz$-plane, as illustrated in Fig.\ \ref{fig_config_z11}. Let the inclination and azimuth angles $(\theta, \phi)$ denote the direction of the vector ${\bf J}_{4}$. From Fig.\ \ref{fig_config_z11}, we see that $\phi$ is the angle between the $({\bf J}_{1}, {\bf J}_5)$ plane and the $({\bf J}_{1}, {\bf J}_{4})$ plane. We denote this angle by $\phi = \phi_{1}$. The inclination angle $\theta$ is the angle between the vectors ${\bf J}_{1}$ and ${\bf J}_{4}$.

The gauge choices for the spinors at the reference point $z_{11}$ are arbitrary, and they only contribute a phase that can be absorbed into $e^{i \kappa_1}$. To be concrete, since ${\bf J}_{1}$ points in the $z$ direction, we choose the spinor $\tau^a(z_{11})$ to be the $\mu^{\text th}$ standard eigenvector for $S_z$, that is,

\begin{equation}
\tau_\alpha^a ( z_{11} ) = \delta_{\alpha \mu}  \, . 
\end{equation}
For the spinor $\tau^b(z_{11})$, we choose it to be an eigenvector of ${\bf J}_{4} \cdot {\bf S}$ in the north standard gauge, that is,

\begin{equation}
\tau_\alpha^b(z_{11}) = e^{i (\alpha - \nu) \phi_{1}} \, d^s_{\nu \alpha } (\theta) \, .
\end{equation}
Taking the spinor inner product, we obtain  

\begin{equation}
\label{eq_spinor_prod_11}
(\tau^b(z_{11}))^\dagger (\tau^a(z_{11}))  = e^{- i (\mu - \nu) \phi_{1}} \, d^s_{\nu \mu} (\theta) \, .
\end{equation}
To evaluate the other spinor product at $z_{12}$, we need to find the rotation matrices $U_a^{(1)}$ and $U_b^{(1)}$, which are generated from paths $\gamma_a$ and $\gamma_b$ from $z_{11}$ to $z_{12}$ along ${\mathcal L}_a^{6js}$ and ${\mathcal L}_b^{6js}$, respectively.

We choose the path $\gamma_a$ to be the path from $p$ to $q$ generated by the ${\bf J}_{34}^2$-flow and the ${\bf J}_{346}^2$-flow, which are illustrated in Fig.\ \ref{fig_loop_large_space} in the large phase space, and in part (a) of Fig.\ \ref{fig_loop_small_space} in the angular momentum space. This path contains no flow generated by the total angular momentum, so 

\begin{equation}
U_a^{(1)} = 1 \, . 
\end{equation}
we choose the path $\gamma_b$ to be the inverse of the path from $q$ back to $p$ along ${\mathcal L}_b^{6js}$ in Fig.\ \ref{fig_loop_large_space}, which contains the overall rotations 

\begin{equation}
U_b^{(1)} = U( {\bf \hat{j}}_{346}, 2 \phi_{346})  U( {\bf \hat{j}}_6, 2 \phi_6) \, . 
\end{equation}
 
\begin{figure}[tbhp]
\begin{center}
\includegraphics[width=0.30\textwidth]{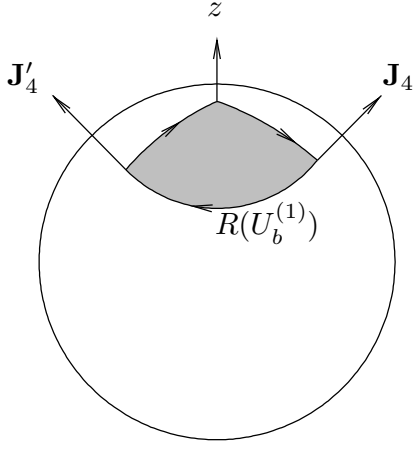}
\caption{The phase difference between two gauge choices can be expressed as an area around a closed loop on the unit sphere.}
\label{fig_spherical_area}
\end{center}
\end{figure}

Because only overall rotations can move the vectors ${\bf J}_4$ and ${\bf J}_6$ along the flows on ${\mathcal L}_b^{6js}$, we can determine this rotation by looking at its effect on ${\bf J}_4$ and ${\bf J}_6$. The effect of the rotation on ${\bf J}_4$ and ${\bf J}_6$ is to reflect  them across the 1-5-346 triangle. In the particular frame that we chose, the rotation $U_b^{(1)}$ effectively moves ${\bf J}_{4}$ to its mirror image ${\bf J}_{4}'$ across the 1-5-346 triangle in the $xz$-plane, which has the direction given by  $(- \phi_{1}, \theta)$. Thus $U_b \, \tau^b(z_{11})$ is an eigenvector of ${\bf J}_{4}' \cdot {\bf S}$, and is up to a phase equal to the eigenvector of ${\bf J}_{4}' \cdot {\bf S}$ in the north standard gauge. Thus, we have

\begin{equation}
[U_b^{(1)} \, \tau^b(z_{11})]_\alpha = e^{i \nu H_{4}}  \, e^{- i (\alpha - \nu) \phi_{1}} \, d^s_{\nu \alpha } (\theta)  \, ,
\end{equation}
where $H_{4}$ is a holonomy phase factor equal to the area of a spherical triangle on a unit sphere. See Fig.\ \ref{fig_spherical_area}. Therefore, the spinor product at the intersection $I_{12}$ is 

\begin{equation}
\label{eq_spinor_prod_12}
( U_b^{(1)} \, \tau^b(z_{11}))^\dagger (U_a \tau^a(z_{11})) = e^{i \nu H_{4}}  \, e^{i (\mu - \nu) \phi_{1}} \, d^s_{\nu \mu} (\theta)  \, .
\end{equation}

Let us denote the first term in Eq.\  (\ref{eq_general_formula_1}) by $T_1$. Substituting the spinor inner products Eq.\  (\ref{eq_spinor_prod_11}) and Eq.\  (\ref{eq_spinor_prod_12}) into Eq.\  (\ref{eq_general_formula_1}), we find that $T_1$ is given by

\begin{eqnarray}
\label{eq_12j_formula_3a}
T_1
&=& e^{i \kappa_1}  \frac{\sqrt{J_{34} J_{346} J_{13} J_{135}}}{ \pi \sqrt{| V_{135} V_{346}  |} }  \, d^s_{\nu \mu} (\theta)    \\  \nonumber
 && \cos \left[ \frac{S^{(1)}}{2}  + \frac{\pi}{2} + \mu \phi_{1} + \nu \left( \frac{H_{4}}{2} - \phi_{1} \right)  \right]  \, . 
\end{eqnarray}

\begin{figure}[tbhp]
\begin{center}
\includegraphics[width=0.30\textwidth]{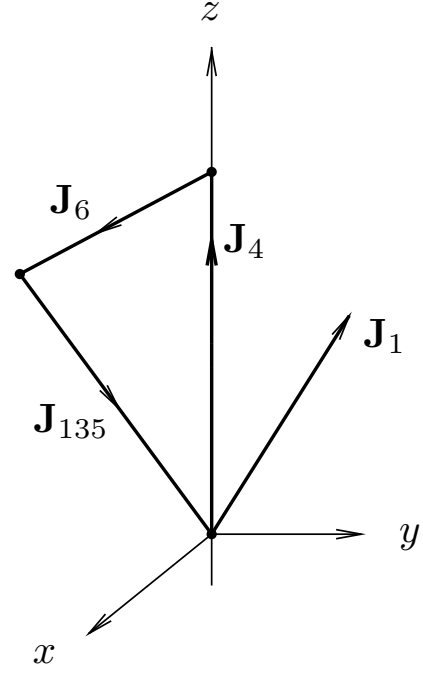}
\caption{A different choice of vector configuration at $z_{11}$ in $I_{11}$.}
\label{fig_config_z11_2}
\end{center}
\end{figure}

Using a different choice of the reference point in Fig.\ (\ref{fig_config_z11_2}) and a different set of paths, we can derive an alternative expression for the inner product, and eliminate the term $H_{4}$. Let us choose a new reference point $z_{11}$ to correspond to an orientation in which ${\bf J}_{4}$ is along the $z$-axis, and ${\bf J}_6$ lies in the $x$-$z$ plane. Through essentially the same arguments as above, we find 

\begin{eqnarray}
\label{eq_12j_formula_3b}
T_1
&=& e^{i \kappa_1}  \frac{\sqrt{J_{34} J_{346} J_{13} J_{135}}}{ \pi \sqrt{| V_{135} V_{346}  |} }  \, d^s_{\nu \mu} (\theta)    \\  \nonumber
 &&  \cos \left[ \frac{S^{(1)}}{2}  + \frac{\pi}{2} + \nu \phi_{4} + \mu \left( \frac{H_{1}}{2} - \phi_{4} \right)  \right]  \, . 
\end{eqnarray}
Here $H_{1}$ is another holonomy for the ${\bf J}_{1}$ vector, and the angle $\phi_{4}$ is the angle between the (${\bf J}_{4}$, ${\bf J}_{6}$) plane and (${\bf J}_{4}$, ${\bf J}_1$) plane.  Because the quantities $S^{(1)},  \phi_{1}, \phi_{4}, H_{1}, H_{4}$ depend only on the geometry of the vector configuration, and are independent of $\mu$ and $\nu$, we conclude that the argument in the cosine must be linear in $\mu$ and $\nu$. Equating the two arguments of the cosine in Eq.\  (\ref{eq_12j_formula_3a}) and in Eq.\  (\ref{eq_12j_formula_3b}), we find that this linear term is $( \mu \phi_{1} + \nu \phi_{4} )$. We find

\begin{eqnarray}
\label{eq_main_formula_wo_phase1}
T_1
&=& e^{i \kappa_1}  \frac{\sqrt{J_{34} J_{346} J_{13} J_{135}}}{ \pi \sqrt{| V_{135} V_{346}  |} }  \, d^s_{\nu \mu} (\theta)    \\   \nonumber
&&   \cos \left[ \frac{S^{(1)}}{2} + \mu \phi_{1}^{(1)} + \nu \phi_{4}^{(1)}  + \frac{\pi}{2}   \right]  \, . 
\end{eqnarray}
Here we have put back the indices $(1)$ to indicate that we are using the vector configuration in which the two tetrahedra are on opposite sides of the 3-346-135 triangle. Through an analogous calculation, we find

\begin{eqnarray}
\label{eq_main_formula_wo_phase2}
T_2
&=& e^{i \kappa_2 } \frac{\sqrt{J_{34} J_{346} J_{13} J_{135}}}{ \pi \sqrt{| V_{135} V_{346}  |} }  \, d^s_{\nu \mu} (\theta)    \\  \nonumber
&&   \cos \left[ \frac{S^{(2)}}{2} - \mu \phi_{1}^{(2)} + \nu \phi_{4}^{(2)}  \right]  \, . 
\end{eqnarray}  \\
Here the indices $(2)$ indicate that we are using the vector configuration in which the two tetrahedra are on same side of the 3-346-135 triangle.

\section{\label{sec_final_formula}An Asymptotic Formula for the $12j$-Symbol}

From the definition Eq.\  (\ref{eq_12j_definition}), we see that the factor $( [j_{34}][j_{346}] [j_{13}][j_{135}] )^{1/2}$ in the denominator of Eq.\  (\ref{eq_12j_definition}) partially cancels out the factor $( J_{34} J_{346} J_{13} J_{135} )^{1/2}$ from $T_1$ and $T_2$ in Eq.\  (\ref{eq_main_formula_wo_phase1}) and Eq.\  (\ref{eq_main_formula_wo_phase2}), respectively, leaving a constant factor of $1/4$. Because the $12j$ symbol is a real number, the relative phase between $e^{i \kappa_1}$ and $e^{i \kappa_2}$ must be $\pm 1$. Through numerical experimentation, we found it to be $(-1)^{2 s_2}$. We use the limiting case of $j_2 = s = 0$ from Eq.\ (\ref{eq_jahn_A8}) to determine the overall phase convention. This determines most of the overall phase. The rest can be fixed through numerical experimentation. Putting the pieces together, we obtain an asymptotic formula for the $12j$ symbol with one small quantum number:

\begin{widetext}
\begin{eqnarray}
\label{eq_main_formula_12j}
 \left\{
  \begin{array}{cccc}
    j_1 & s & j_{12} & j_{125} \\ 
    j_3 & j_4 & j_{34} &  j_{135}  \\ 
    j_{13} & j_{24} & j_5 & j_6  \\
  \end{array} 
  \right\}   
=   \frac{(-1)^{j_1 +2 j_3 + j_4 + j_{346} + j_{135} + j_6 + s + \mu }}{4 \pi \sqrt{| V_{135} V_{346}  |} }  \, 
 &&  \{  d^{s}_{\nu \mu} (\theta^{(1)})   \cos \left[ S_1 + S_2  + \mu \phi_{1}^{(1)} + \nu \phi_{4}^{(1)}  + \frac{\pi}{2} \right] \\  \nonumber
&&  \quad  + (-1)^{2 s} \, d^{s}_{\nu \mu} (\theta^{(2)}) \cos \left[ S_1 - S_2 - \mu \phi_{1}^{(2)} + \nu \phi_{4}^{(2)}  \right]    \}
\end{eqnarray}
\end{widetext}

\begin{figure}[tbhp]
\begin{center}
\includegraphics[width=0.50\textwidth]{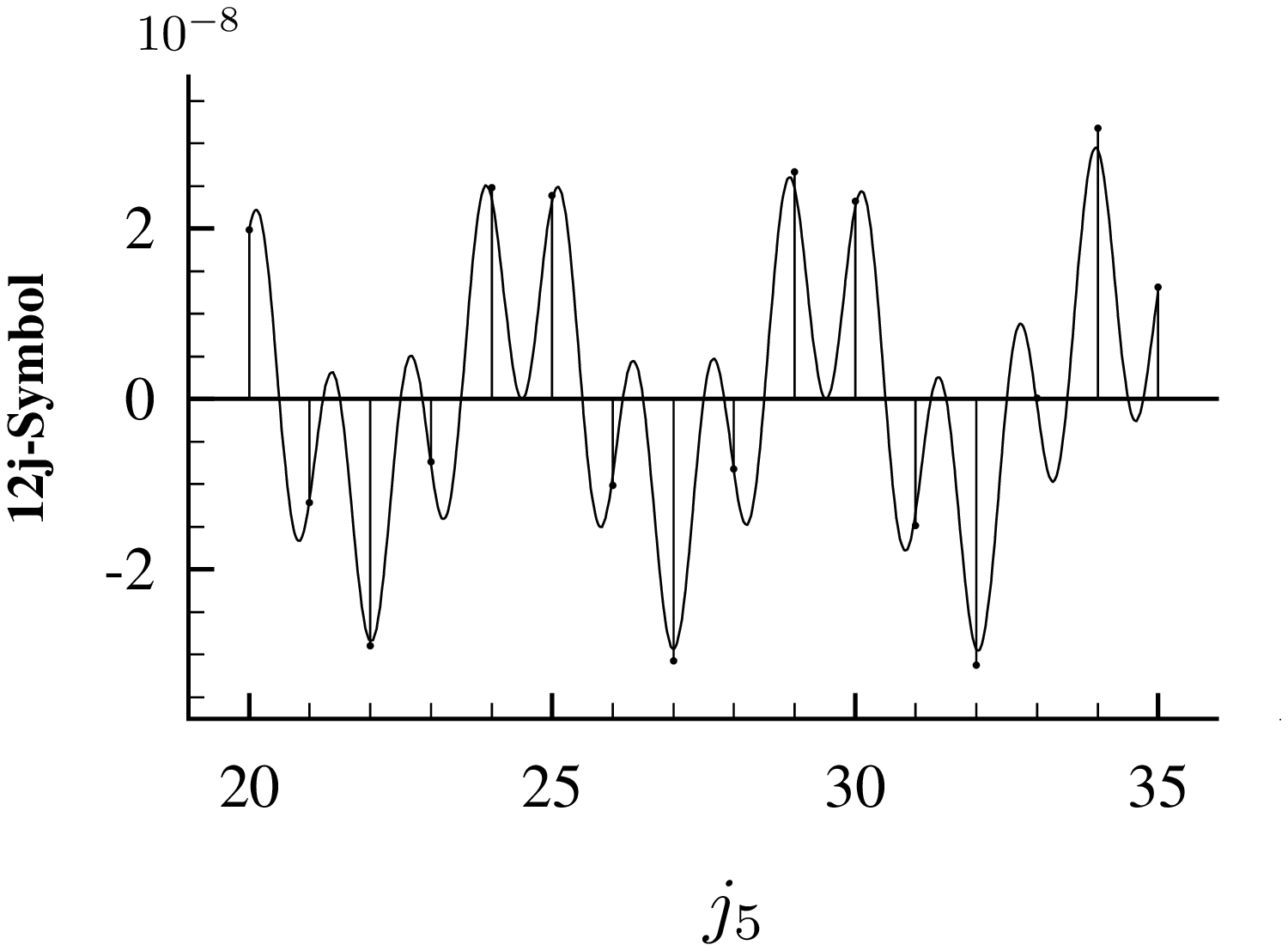}
\caption{Comparison of the exact $12j$ symbol (vertical sticks and dots) and the asymptotic formula (\ref{eq_main_formula_12j}) in the classically allowed region away from the caustics, for the values of $j$'s shown in Eq.\  (\ref{eq_12j_values_1_11_case1}). }
\label{fig_12j_plot_1_case1}
\end{center}
\end{figure}

\begin{figure}[tbhp]
\begin{center}
\includegraphics[width=0.50\textwidth]{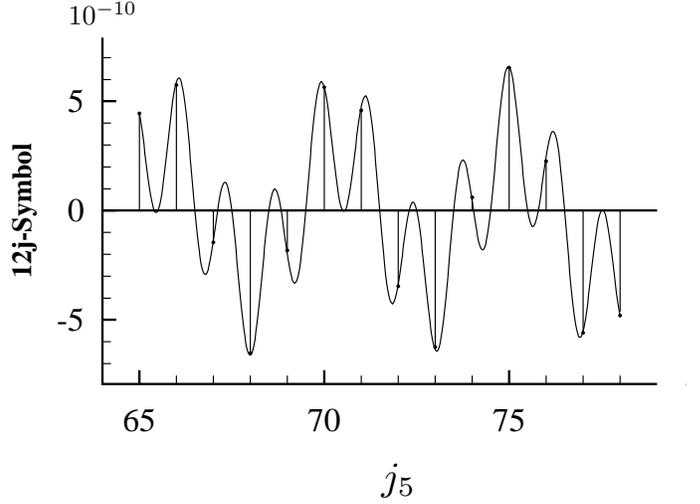}
\caption{Comparison of the exact $12j$ symbol (vertical sticks and dots) and the asymptotic formula (\ref{eq_main_formula_12j}) in the classically allowed region away from the caustics, for the values of $j$'s shown in Eq.\  (\ref{eq_12j_values_1_11}). }
\label{fig_12j_plot_1}
\end{center}
\end{figure}

Here, the indices on the $d$-matrix are given by $\mu = j_{12}-j_{1}$ and $\nu = j_{24}-j_{4}$. They are of the same order as the small parameter $s$. The two Ponzano-Regge phases $S_1$ and $S_2$ are defined in Eq.\ (\ref{eq_S_PR_346}) and Eq.\ (\ref{eq_S_PR_135}), respectively. The quantity $V_{ijk}$ is defined by

\begin{equation}
V_{ijk} = {\bf J}_i \cdot ({\bf J}_j \times {\bf J}_k) \, . 
\end{equation}

The angle $\theta$ is the angle between the vectors ${\bf J}_{12}$ and ${\bf J}_{13}$. The angles $\phi_1, \phi_4, \theta$ are given by the following equations

\begin{eqnarray}
\label{eq_phi_11_def}
\phi_{1} &=&   \pi - \cos^{-1} \left( \frac{ ({\bf J}_{1} \times {\bf J}_{4} ) \cdot ({\bf J}_{1} \times {\bf J}_{5} ) }{ | {\bf J}_{1} \times {\bf J}_{4} | \,  | {\bf J}_{1} \times {\bf J}_{5}  |} \right)   \, ,    \\
\label{eq_phi_41_def}
\phi_{4} &=&  \pi -  \cos^{-1} \left(  \frac{ ({\bf J}_{4} \times {\bf J}_{1} ) \cdot ({\bf J}_{4} \times {\bf J}_{6} ) }{ | {\bf J}_{4} \times {\bf J}_{1} | \,  | {\bf J}_{4} \times {\bf J}_{6}  |}  \right)   \, ,   \\
\label{eq_theta_1_def}
\theta &=& \cos^{-1} \left( \frac{ {\bf J}_{1} \cdot {\bf J}_{4} }{J_{1} J_{4} }  \right)  \, . 
\end{eqnarray}

These angles are calculated using the vector configuration at a point in $I_{21}$, which is illustrated in Fig.\ \ref{fig_6js_config_I_21}.

We illustrate the accuracy of the approximation Eq.\  (\ref{eq_main_formula_12j}) by plotting it against the exact $12j$ symbol in the classically allowed region for the following values of the $j$'s:  

\begin{equation}
\label{eq_12j_values_1_11_case1}
\left\{
  \begin{array}{cccc}
    j_1 & s_2 & j_{12} & j_{125} \\ 
    j_3 & j_4 & j_{34} &  j_{135}  \\ 
    j_{13} & j_{24} & j_5 & j_6  \\
  \end{array} 
  \right\}   
=
	\left\{
  \begin{array}{rrrr}
    35 & 1 & 34 & 39 \\ 
    36 & 28 & 38 & 31 \\ 
    27 & 29 & j_5 & 36 \\
  \end{array} 
  \right\} \, . 
\end{equation}
The result is shown in Fig.\ \ref{fig_12j_plot_1_case1}.

Since the asymptotic formula (\ref{eq_main_formula_12j}) should become more accurate as the values of the $j$'s get larger, we plot the formula against the exact $12j$ symbol for another example,  

\begin{equation}
\label{eq_12j_values_1_11}
\left\{
  \begin{array}{cccc}
    j_1 & s_2 & j_{12} & j_{125} \\ 
    j_3 & j_4 & j_{34} &  j_{135}  \\ 
    j_{13} & j_{24} & j_5 & j_6  \\
  \end{array} 
  \right\}   
=
	\left\{
  \begin{array}{rrrr}
    177/2 & 5/2 & 88 & 89 \\ 
    181/2 & 141/2 & 87 & 77 \\ 
    75 & 73 & j_5 & 91 \\
  \end{array} 
  \right\} \, ,
\end{equation}
in the classically allowed region away from the caustic in Fig.\ \ref{fig_12j_plot_1}. These values of the $j$'s are roughly $2.5$ times those in Eq.\  (\ref{eq_12j_values_1_11_case1}). Again, the agreement has improved.

\section{Conclusions}

In this paper, we have derived the second asymptotic formula for the Wigner $12j$ symbol with one small angular momentum. Eq.\ (\ref{eq_main_formula_12j}) here and Eq.\ (80) in \cite{yu2011b} together cover all the different placements of the small angular momentum among the 12 parameters of the $12j$ symbol. Although the two asymptotic formulas of the $12j$ symbol are similar, in that they are expressed in terms of the asymptotic phases of lower $3nj$-symbols, Eq.\ (\ref{eq_main_formula_12j}) in this paper is in many ways simpler. The relationship between the $6j$ symbol and the geometry of a tetrahedron is well known. The construction of the vectors of the tetrahedra is simpler than that for the vector diagram of a $9j$ symbol in \cite{yu2011b}. In any case, the two formulas are valid even when we take $s$ to be large, as long as the other $11$ angular momenta $j$ are much larger relative to $s$, that is, as long as $1 << s << j$. 

Currently, the asymptotic formula for the $12j$ symbol when all the angular momenta are large is still unknown. Such a formula must reduce to Eq.\ (\ref{eq_main_formula_12j}) here or Eq.\ (80) in \cite{yu2011b} in the limit $1 << s << j$. Therefore, the work in this paper may eventually help us find the asymptotic formula of the $12j$ symbol when all $j$ are large. 

After the completion of this manuscript, I was informed by the authors of \cite{bonzom2011} that they have found an independent derivation for the asymptotic formula of the $9j$ symbol in \cite{yu2011a}.

\section*{Adknowledgement}

The author would like to thank Luyen Pham for her help in editing the manuscript. 


\bibliography{SLQN12jB}

\end{document}